\documentclass[aps, prd,  superscriptaddress, showpacs, longbibliography, floatfix, nofootinbib]{revtex4-2}

\usepackage{bm, amsmath, amssymb, relsize, amsfonts, mathrsfs, multirow, braket, siunitx, color, booktabs, arydshln, tabularx}
\usepackage{graphicx}
\graphicspath{{./figures/}}

\DeclareSIUnit \parsec {pc}

\usepackage{layouts}
\usepackage[dvipsnames]{xcolor}
\usepackage[unicode]{hyperref}
\hypersetup{
  colorlinks=true,
  citecolor=RoyalBlue,
  linkcolor=blue,
  urlcolor=blue
}

\usepackage{array}
\usepackage{cancel}
\usepackage{orcidlink}
\usepackage{dcolumn}
\usepackage{bm}
\usepackage{paralist}
\usepackage{epsfig}
\usepackage{placeins}
\usepackage{mathrsfs}
\usepackage{comment}
\usepackage{color, colortbl}
\usepackage[inline]{enumitem}
\usepackage{soul}
\usepackage{adjustbox}
\usepackage{amssymb}
\usepackage{caption}
\usepackage{subcaption}
\usepackage{float}
\usepackage{cancel}
\usepackage{verbatim}
\usepackage{amsmath}
\usepackage[toc,page]{appendix}
\usepackage{graphicx}
\usepackage{ragged2e}

\DeclareMathAlphabet{\mathpzc}{OT1}{pzc}{m}{it}
	
\definecolor{LightCyan}{rgb}{0.88,1,1}
\definecolor{lightgray}{gray}{0.9}

\def \IITGn     {Department of Physics, Indian Institute of Technology Gandhinagar, Gujarat 382355, India.\vspace*{4pt}}
\def \HRI     {Regional Centre for Accelerator-based Particle Physics, Harish-Chandra Research Institute, HBNI, Chhatnag Road, Jhunsi, Prayagraj 211019, India.\vspace*{4pt}}

\begin{document}
\title{LHC Signatures of Neutral Scalar Cascades in the $Z_3$ symmetric 3HDM}

\author{\textsc{Baradhwaj Coleppa}\orcidlink{0000-0002-8761-3138}}
\email{baradhwaj@iitgn.ac.in}
\affiliation{\IITGn}

\author{\textsc{Akshat Khanna}\vspace*{7pt}\orcidlink{0000-0002-2322-5929}}
\email{khanna\textunderscore akshat@iitgn.ac.in}
\affiliation{\IITGn}

\author{\textsc{Santosh Kumar Rai}\vspace*{7pt}\orcidlink{0000-0002-4130-6992}\vspace*{7pt}}
\email{skrai@hri.res.in}
\affiliation{\HRI}

\author{\textsc{Agnivo Sarkar}\vspace*{7pt}\orcidlink{0000-0001-9596-1936}\vspace*{7pt}}
\email{agnivosarkar@hri.res.in}
\affiliation{\HRI}

\begin{flushright}
    HRI-RECAPP-2026-01
\end{flushright}

\begin{abstract}
Extending the scalar sector is one of the standard approaches to exploring scenarios beyond the Standard Model. In this work, we examine the collider phenomenology of the Three Higgs Doublet Model (3HDM) in the Type-Z or the democratic Yukawa interaction setup at the LHC. The scalar spectrum of the 3HDM includes three CP-even scalars, two CP-odd scalars, and four charged Higgs bosons. Focusing on cascade decay topologies, we investigate the collider signatures of the neutral scalars through the process $pp \rightarrow SV$, where $S$ is a neutral scalar and $V$ is a vector boson. We perform a cross-section analysis across multiple benchmark points that satisfy both theoretical and experimental constraints, considering two mass hierarchy scenarios: (i) Regular Hierarchy, where the SM-like Higgs is the lightest CP-even scalar, and (ii) Medial Hierarchy, featuring one Higgs boson lighter than the SM Higgs and one heavier. For both scenarios, we study the specific process $pp \rightarrow A \rightarrow HZ \rightarrow b \bar{b} l^+l^-$, performing a cut and count analysis at $\sqrt{s}=14$ TeV. Our results demonstrate that while the Medial Hierarchy scenario allows discovery-level sensitivity for both the CP-even and CP-odd scalars, achieving the same sensitivity in the Regular Hierarchy setup necessitates substantially higher luminosity.
\end{abstract}
\maketitle
\section{Introduction}
\label{sec:intro}

The Standard Model (SM) of particle physics \cite{Salam:1968rm,Weinberg:1967tq,Glashow:1961tr} is a remarkably successful $SU(3)_c\times SU(2)_L\times U(1)_Y$ gauge theory that has consistently withstood experimental scrutiny. The discovery of the $125$ GeV Higgs boson at the Large Hadron Collider (LHC) in 2012 \cite{Akeroyd_2021,CMS:2012qbp} marked the completion of the SM particle spectrum. However, despite this monumental achievement, several properties of the Higgs boson remain to be precisely measured, and its role in the broader context of particle physics is still not fully understood. These open questions, along with the inability of the SM to address phenomena such as dark matter, neutrino masses, and the baryon asymmetry of the universe, motivate the pursuit of theoretical frameworks that extend beyond the Standard Model (BSM).

One promising direction for BSM exploration involves extending the scalar sector of the SM. The Two Higgs Doublet Model (2HDM) and its various realizations \cite{Branco:2011iw,Coleppa:2013dya,Chang:2012ve, Grinstein:2013npa,Drozd:2014yla,Bhattacharya:2023qfs,Muhlleitner:2016mzt,Keus:2017ioh,Bhattacharyya:2015nca,Cao:2009as} serve as canonical examples and have been widely studied for their rich phenomenological implications \cite{Ko:2013zsa,Crivellin:2013wna,PhysRevD.106.075003,PhysRevD.109.055016,PhysRevD.100.035031,Logan:2010ag}. Likewise, the Three Higgs Doublet Model (3HDM) has also been an active area of research \cite{Cordero:2017owj,Ivanov:2012ry,Akeroyd:2021fpf,Boto:2025ovp,Ivanov:2014doa,Keus:2013hya}. As its name suggests, the 3HDM extends the scalar sector by introducing two additional $SU(2)_L$ scalar doublets, leading to a more intricate scalar spectrum and enhanced potential for novel signatures at colliders. Furthermore, the 3HDM allows for a variety of Yukawa coupling structures, offering different mechanisms for fermion mass generation. Several studies \cite{Batra:2025amk,Keus:2013hya,Coleppa:2025qst,PhysRevD.108.015020,PhysRevD.111.075009,PhysRevD.100.015008,Kalinowski:2021lvw} have examined the theoretical consistency and phenomenological viability of this model under various symmetry assumptions and coupling configurations. 

A distinctive feature of extended Higgs sectors is the presence of the CP-odd scalar, $A$, which is absent in the SM and therefore serves as a sensitive probe of new physics. In particular, the decay channel $A \rightarrow HZ$, where $H$ denotes a CP-even scalar, offers a clean experimental signature, when followed by a leptonic decay of the $Z$ boson and the scalar decaying to a $b$-quark pair. This process has already been the subject of dedicated searches at the ATLAS and CMS experiments \cite{CMS:2015flt,CMS-PAS-HIG-14-011,CMS:2019qcx}, highlighting its phenomenlogical importance. Importantly, the $AHZ$ coupling originates from scalar mixing effects in multi higgs doublet models, making it a direct probe of the underlying Higgs structure. 
Heavy neutral Higgs bosons in extended scalar sectors can be searched for at colliders through several complementary channels. In scenarios where additional Higgs doublets or triplets participate in EWSB, the heavy scalars typically couple to the SM gauge bosons, rendering decays such as $H \to VV$ a useful probe. In addition, the Higgses can decay to fermion pairs, particularly $b\bar{b}$ and $t\bar{t}$ (if kinematically accessible) - however typically these involve a formidable QCD background and are challenging. A particularly promising third avenue arises when heavy Higgs bosons decay into lighter new-physics states, leading to cascade decay topologies with characteristic kinematic features. In the class of 3HDM considered in this work, imposing the alignment limit implies that the couplings of the non SM-like neutral Higgs bosons to SM gauge bosons vanish identically,\footnote{This feature is also present in the 2HDM and certainly not unique to the 3HDM framework.} thereby strongly suppressing direct $H \to VV$ signatures. Consequently, cascade decays involving transitions between scalar states accompanied by a SM gauge boson can become the phenomenologically relevant channels. These processes not only retain sizable signal rates but also benefit from comparatively cleaner experimental signatures and more controllable SM backgrounds, making them well suited for collider-based investigations. 

In this work, we focus on a $Z_3$-symmetric realization of the 3HDM with a Type-Z or a democratic Yukawa structure. We investigate the collider phenomenology of this model by analyzing a specific class of cascade decays in two distinct scalar mass hierarchies: (i) Regular Hierarchy, in which the SM-like Higgs is the lightest CP-even scalar, and (ii) Medial Hierarchy, where it occupies an intermediate position in the mass spectrum. For each scenario, we select a benchmark point that satisfies all relevant theoretical constraints (such as vacuum stability and perturbative unitarity) and experimental bounds (from Higgs signal strengths, direct searches, and flavor observables). Our phenomenological analysis focuses on the process $pp \rightarrow A \rightarrow HZ$, where the $H$ subsequently decays to a pair of bottom quarks, while the $Z$ boson decays leptonically. A cut-and-count strategy is employed with the LHC center-of-mass energy fixed $\sqrt{s}=14$ TeV and with an integrated luminosity of $200 \ \textrm{fb}^{-1}$. We find that a discovery-level significance of $5\sigma$ can be achieved for both the CP-even and CP-odd scalars in the Medial Hierarchy scenario, whereas attaining a comparable significance for the CP-odd scalar in the Regular Hierarchy case requires substantially higher integrated luminosity.

The structure of this paper is as follows: in Section~\ref{sec:model}, we describe the scalar content of the specific 3HDM under study and outline the key parameters and symmetries that define this $Z_3$-symmetric version. The Yukawa sector with Type-Z structure is then introduced, followed by a discussion of the relevant scalar couplings and constraints. Section~\ref{sec:alignment} discusses the alignment limit conditions under which one of the CP-even scalars mimics the SM Higgs boson. The collider phenomenology for both mass hierarchies is presented in Section~\ref{sec:pheno}, including cross-section calculations and significance estimates. We conclude with a summary of our results and outlook in Section~\ref{sec:conclusions}.

\section{3HDM: Model}
\label{sec:model}
In this section, we provide a comprehensive account of the model construction. We begin with an overview of the scalar sector, followed by a discussion of the Yukawa interactions. Key couplings are then summarized in tabular form. Finally, we detail the theoretical and experimental constraints imposed on the parameter space of the model.
\subsection{Scalar Sector}
The scalar sector of the 3HDM features three $SU(2)_L$ doublets which are represented as 
\begin{equation}
		\Phi_k = \begin{pmatrix}
			\phi_k^+ \\ \frac{v_k+p_k+in_k}{\sqrt{2}}
		\end{pmatrix},
  \label{eq:phi_form}
\end{equation}
where $k=1,2,3$. We write the most general scalar potential that is invariant under the gauge group $SU(2)_L\times U(1)_Y$ and incorporates a $Z_3$ symmetry:
\begin{equation}
     	\label{eq:scalarpot}
     	\begin{split}
     		V & = m_{11}^2(\Phi_1^\dagger\Phi_1) + m_{22}^2(\Phi_2^\dagger\Phi_2) + m_{33}^2(\Phi_3^\dagger\Phi_3)
     		\\ & + \lambda_1(\Phi_1^\dagger\Phi_1)^2 + \lambda_2(\Phi_2^\dagger\Phi_2)^2 + \lambda_3(\Phi_3^\dagger\Phi_3)^2 \\ & + \lambda_{4}(\Phi_1^\dagger\Phi_1)(\Phi_2^\dagger\Phi_2) + \lambda_{5}(\Phi_1^\dagger\Phi_1)(\Phi_3^\dagger\Phi_3) + \lambda_{6}(\Phi_2^\dagger\Phi_2)(\Phi_3^\dagger\Phi_3) \\ & + \lambda_{7}(\Phi_1^\dagger\Phi_2)(\Phi_2^\dagger\Phi_1) + \lambda_{8}(\Phi_1^\dagger\Phi_3)(\Phi_3^\dagger\Phi_1) +  \lambda_{9}(\Phi_2^\dagger\Phi_3)(\Phi_3^\dagger\Phi_2) \\ & + [\lambda_{10}(\Phi_1^\dagger\Phi_2)(\Phi_1^\dagger\Phi_3) + \lambda_{11}(\Phi_1^\dagger\Phi_2)(\Phi_3^\dagger\Phi_2) + \lambda_{12}(\Phi_1^\dagger\Phi_3)(\Phi_2^\dagger\Phi_3) + h.c.]. \\ &
     	\end{split} 
\end{equation}
Under this additional $Z_3$ symmetry, the Higgs fields transform as
\begin{equation}
     	\label{eq:phitrans}
     	\Phi_1 \rightarrow \omega \Phi_1, \; \; \Phi_2 \rightarrow \omega^2 \Phi_2, \;\textrm{and} \;\Phi_3 \rightarrow \Phi_3,
\end{equation}
where $\omega = e^{2\pi i/3}$ are the cube roots of unity. In general, the parameters $\lambda_{1,2,...,9}$ in the scalar potential are real, as required by the Hermiticity of the Lagrangian, whereas $\lambda_{10},\lambda_{11}$ and $\lambda_{12}$ may, in principle, be complex. To avoid mixing between CP-even and CP-odd scalar states, we consider a CP-conserving scenario by setting all potentially complex parameters in the potential to zero. We start by examining the relevant mass terms for the CP-even Higgs bosons, which can be directly derived from the scalar potential given in Eqn.~\ref{eq:scalarpot}, and schematically expressed as
\begin{equation*}
		V_p^{mass} \supset \begin{pmatrix}
			p_1 & p_2 & p_3
		\end{pmatrix} \frac{\mathcal{M}^2_S}{2} \begin{pmatrix}
			p_1 \\ p_2 \\ p_3 
		\end{pmatrix}.
\end{equation*}
The real symmetric mass matrix $\mathcal{M}^2_S$ can be diagonalized by an orthogonal transformation by a matrix $O_\alpha$ defined as
\begin{equation}
    \label{eq:matalphtrans}
    O_\alpha = \begin{pmatrix}
        c_{\alpha 1} c_{\alpha 2} & c_{\alpha 2} s_{\alpha 1} & s_{\alpha 2} \\ -c_{\alpha 3} s_{\alpha 1} - s_{\alpha 3} s_{\alpha 2} c_{\alpha 1} & c_{\alpha 3} c_{\alpha 1} - s_{\alpha 3} s_{\alpha 2} s_{\alpha 1} & s_{\alpha 3} c_{\alpha 2} \\ s_{\alpha 3} s_{\alpha 1} - c_{\alpha 3} s_{\alpha 2} c_{\alpha 1} & -s_{\alpha 3} c_{\alpha 1} - c_{\alpha 3} s_{\alpha 2} s_{\alpha 1} & c_{\alpha 3} c_{\alpha 2} 
    \end{pmatrix}.
\end{equation}
The diagonalization condition implies
\begin{equation}
    O_\alpha.\mathcal{M}^2_S.O_\alpha^T = \begin{pmatrix}
        m_{H 1}^2 & 0 & 0 \\ 0 & m_{H 2}^2 & 0 \\ 0 & 0 & m_{H 3}^2
    \end{pmatrix}.
\end{equation}
In this work, to facilitate phenomenological exploration, we consider the scalar masses and mixing angles as independent parameters, and compute the corresponding couplings $\lambda_i's$ in terms of them by inverting the above relation \cite{Batra:2025amk}. 
The gauge eigenstates can be written down in terms of the mass eigenstates using the transformation matrix in a straightforward manner: 
\begin{equation}
    \begin{split}
        H_1 & = c_{\alpha_2} c_{\alpha_1} p_1 + c_{\alpha_2} s_{\alpha_1} p_2 + s_{\alpha_2}p_3, \\
        H_2 & = -(c_{\alpha_3} s_{\alpha_1} + s_{\alpha_3} s_{\alpha_2} c_{\alpha_1})p_1 + (c_{\alpha_3} c_{\alpha_1} - s_{\alpha_3} s_{\alpha_2} s_{\alpha_1})p_2+(s_{\alpha_3} c_{\alpha_2})p_3,\,\textrm{and} \\
        H_3 & =  (s_{\alpha_3} s_{\alpha_1} - c_{\alpha_3} s_{\alpha_2} c_{\alpha_1})p_1 -  (s_{\alpha_3} c_{\alpha_1} + c_{\alpha_3} s_{\alpha_2} s_{\alpha_1})p_2 + (c_{\alpha_3} c_{\alpha_2}) p_3.
    \end{split}
\end{equation}
The mass terms for the charged Higgses can similarly be extracted from the scalar potential given in Eqn.~\ref{eq:scalarpot}:
\begin{equation*}
     	V_C^{mass} \supset \begin{pmatrix}
     		\phi_1^- & \phi_2^- & \phi_3^-
     	\end{pmatrix} \mathcal{M}^2_{\phi^{\pm}} \begin{pmatrix}
     		\phi_1^+ \\ \phi_2^+ \\ \phi_3^+ 
     	\end{pmatrix},
\end{equation*}
where, $\mathcal{M}^2_{\phi^{\pm}}$ is the $3 \times 3$ charged Higgs mass matrix. To diagonalize this, we first employ a similarity transformation using the matrix $O_\beta$: 
\begin{equation*}
    (B_C)^2 = O_\beta.\mathcal{M}^2_{\phi^{\pm}}.O_\beta^T,
\end{equation*}
where
\begin{eqnarray}
    \label{eq:betatrans}
    O_\beta &=
     & \begin{pmatrix}
        c_{\beta 2}c_{\beta 1} & c_{\beta 2}s_{\beta 1} & s_{\beta 2} \\ -s_{\beta 1} & c_{\beta 1} & 0 \\ -c_{\beta 1}s_{\beta 2} & -s_{\beta 1}s_{\beta 2} & c_{\beta 2}
    \end{pmatrix}.
\end{eqnarray}
Here $\tan\beta_1=v_2/v_1$ and $\tan\beta_2=v_3/\sqrt{v_1^2+v_2^2}$. The above matrix $B_C$ can be fully diagonalized by a subsequent transformation with matrix $O_{\gamma 2}$:
\begin{equation*}
     	O_{\gamma 2}.(B_C)^2.O_{\gamma 2}^\dagger = \begin{pmatrix}
     		0 & 0 & 0 \\ 0 & m_{H^{\pm}_2}^2 & 0 \\ 0 & 0 & m_{H^{\pm}_3}^2
     	\end{pmatrix}.
\end{equation*}
As in the case of CP-even scalars, analogous relations can be derived connecting the parameters $\lambda_7,\lambda_8,\lambda_9$ to the charged Higgs masses, mixing angles, and the vevs. The gauge eigenstates can hence be represented in terms of mass eigenstates as
\begin{equation}
    \label{chargetransrelat}
    \begin{split}
        G^\pm & = c_{\beta 1}c_{\beta 2} \phi_1^\pm + c_{\beta 2}s_{\beta 1} \phi_2^\pm + s_{\beta 2} \phi_3^\pm, \\
        H_2^\pm & = (-c_{\gamma 2 }s_{\beta 1} + c_{\beta 1}s_{\beta 2}s_{\gamma 2})\phi_1^\pm + (c_{\beta 1}c_{\gamma 2} + s_{\beta 1}s_{\beta 2}s_{\gamma 2})\phi_2^\pm + (-c_{\beta 2}s_{\gamma 2}) \phi_3^\pm,\,\textrm{and} \\
        H_3^\pm & = (-c_{\beta 1}c_{\gamma 2}s_{\beta 2}-s_{\beta 1}s_{\gamma 2})\phi_1^\pm + (-c_{\gamma 2}s_{\beta 1}s_{\beta 2}+c_{\beta 1}s_{\gamma 2})\phi_2^\pm + (c_{\beta 2}c_{\gamma 2}) \phi_3^\pm.
    \end{split}
\end{equation}
Finally, writing the mass terms for the CP-odd Higgs in a similar fashion
\begin{equation*}
        V_n^{mass} \supset \begin{pmatrix}
            n_1 & n_2 & n_3
        \end{pmatrix} \frac{\mathcal{M}^2_n}{2} \begin{pmatrix}
            n_1 \\ n_2 \\ n_3 
        \end{pmatrix}, 
\end{equation*}
we note that the pseudoscalar mass matrix can be diagnolized exactly like in the previous case, i.e., by  performing two consecutive rotations, first by $O_\beta$ and then by $O_{\gamma1}$. Once again, we trade the Lagrangian parameters for the masses, mixing angles, and vevs. The gauge eigenstates can hence be represented in terms of mass eigenstates as
\begin{equation}
    \begin{split}
        G_0 & = (c_{\beta 1}c_{\beta 2}) n_1 + (c_{\beta 2}s_{\beta 1}) n_2 + (s_{\beta 2}) n_3, \\
        A_1 & = (-c_{\gamma 1 }s_{\beta 1} + c_{\beta 1}s_{\beta 2}s_{\gamma 1})n_1 + (c_{\beta 1}c_{\gamma 1} + s_{\beta 1}s_{\beta 2}s_{\gamma 1})n_2 + (-c_{\beta 2}s_{\gamma 1})n_3,\,\textrm{and} \\
        A_2 & = (-c_{\beta 1}c_{\gamma 1}s_{\beta 2}-s_{\beta 1}s_{\gamma 1})n_1 + (-c_{\gamma 1}s_{\beta 1}s_{\beta 2}+c_{\beta 1}s_{\gamma 1})n_2 + (c_{\beta 2}c_{\gamma 1}) n_3.
    \end{split}
\end{equation}
\subsection{Yukawa Sector}
Tree-level Flavor-Changing Neutral Currents (FCNCs) pose a significant challenge to multi-Higgs doublet models due to stringent experimental limits. To avoid such currents, we implement the Natural Flavor Conservation (NFC), which mandates that each class of fermions couples to a single Higgs doublet. This prevents the appearance of FCNCs at tree level by construction. To realize this condition, we adopt the Type-Z Yukawa structure, also referred to as the democratic scenario. In this setup, each Higgs doublet couples to a specific fermion type - up-type quarks, down-type quarks, and charged leptons - ensuring that mass generation occurs independently for each sector. The resulting Yukawa Lagrangian for the model is given by
\begin{equation}
		\label{eq:yukeq}
		\mathcal{L}_{Yukawa} = -[\bar{L}_L \Phi_1 \mathcal{G}_l l_R+\bar{Q}_L \Phi_2 \mathcal{G}_d d_R + \bar{Q}_L \tilde{\Phi}_3 \mathcal{G}_u u_R  + h.c].
\end{equation}
The $\mathcal{G}_f$ are the Yukawa matrices, and in terms of the fermion mass matrices they can be written as
\begin{equation*}
		\mathcal{G}_f = \frac{\sqrt{2} \mathcal{M}_f}{v_i}.
\end{equation*}
We work with a $Z_3$ symmetric potential as given in Eqn.~\ref{eq:phitrans} - for the Yukawa Lagrangian to remain invariant under the same, the right handed fermion fields are required to transform as 
\begin{equation}
		\label{eq:fermtrans}
		d_R  \rightarrow \omega d_R , \; \; \; \; l_R \rightarrow \omega^2 l_R, \; \; \; \; u_R \rightarrow  u_R.
\end{equation}
\subsection{Constraints}
\label{sec:constraints}
To guarantee the stability of the electroweak vacuum, the scalar potential must be bounded from below along all possible directions in field space. This requirement translates into a set of theoretical constraints on the quartic couplings $\lambda_i$'s, ensuring that no field direction leads to a potential unbounded from below. Specifically, the vacuum stability conditions are expressed as
\begin{equation}
    \begin{gathered}
        \lambda_{1} \geq 0, \quad \lambda_{2} \geq 0, \quad \lambda_{3} \geq 0, \\
         \lambda_{4} + 2\sqrt{\lambda_{1}\lambda_{2}} \geq 0, \quad \lambda_{5} + 2\sqrt{\lambda_{1}\lambda_{3}} \geq 0, \quad \lambda_{6} + 2\sqrt{\lambda_{2}\lambda_{3}} \geq 0, \\
         \lambda_{4} + \lambda_{7} + 2\sqrt{\lambda_{1}\lambda_{2}} \geq 0, \quad \lambda_{5} + \lambda_{8} + 2\sqrt{\lambda_{1}\lambda_{3}} \geq 0, \quad \lambda_{6} + \lambda_{9} + 2\sqrt{\lambda_{2}\lambda_{3}} \geq 0.
    \end{gathered}
\end{equation}
Alongside vacuum stability, we also impose unitarity constraints to ensure that the theory remains well-behaved and predictive at high energies. Following the results of \cite{Bento:2022vsb} for the $Z_3$ symmetric 3HDM, these conditions require that the $21$ eigenvalues of the scalar two-to-two scattering matrices satisfy $|\Lambda_i| \leq 8\pi$. Furthermore, the perturbative consistency of the model demands that all quartic couplings remain within the perturbative regime, which we enforce by requiring $\lambda_i \leq |4\pi|$ for all $i$. 
To ensure consistency with electroweak precision measurements, we evaluate the oblique parameters $S$, $T$ and $U$, which encapsulate the impact of new physics on the gauge boson propagators and are known to place stringent constraints on extensions of the SM. 

In addition, we examine constraints on the BSM scalar sector arising from direct searches at the LHC, LEP, and Tevatron, applying the $95 \% $ confidence-level exclusion limits using the \texttt{HiggsBound} framework \cite{Bechtle:2020pkv}, through the \texttt{HiggsTools} interface \cite{Bahl:2022igd}. The compatibility of the 125 GeV Higgs boson in our model with the observed SM-like Higgs signal is tested via a goodness-of-fit analysis using \texttt{HiggsSignal} \cite{Bechtle:2020uwn}. Furthermore, we incorporate flavour physics constraints by considering the most stringent bounds on the branching ratio $\mathcal{BR}$) of the $B\rightarrow X_s \gamma$ computed at next-to-leading order (NLO) in QCD following \cite{Borzumati_1998,Akeroyd_2021,Boto_2021}. The following restriction has been imposed which represents the $3 \sigma$ experimental limit:
\begin{equation*}
    2.87 \times 10^{-4} < \mathcal{BR}(B \rightarrow X_s \gamma) < 3.77 \times 10^{-4}.
\end{equation*}
\section{Alignment Limit in the 3HDM}
\label{sec:alignment}
The observation of a 125 GeV Higgs boson with properties closely resembling those predicted by the SM implies that one of the three CP-even Higgs bosons in the 3HDM must mimic this behavior. In this section, we review the theoretical implications of enforcing the alignment limit, a condition under which one Higgs bosons acquires SM-like couplings. A distinctive feature of the 3HDM is the absence of a fixed mass hierarchy among the CP-even Higgs states -- their ordering depends sensitively on the numerical values assigned to the couplings ($\lambda_i$) and the vev's, allowing for a variety of mass configurations. Consequently, any one of the three CP-even states could, in principle, serve as the SM-like Higgs. To organize our analysis, we consider two\footnote{As demonstrated in our earlier work \cite{Batra:2025amk}, the third scenario, called the Inverted Hierarchy, does not yield any benchmark points that satisfy all theoretical and experimental constraints.} possibilities: \cite{Batra:2025amk}
\begin{itemize}
    \item \textbf{Regular Hierarchy:} here the lightest CP-even scalar is identified as the SM-like Higgs.
    \item \textbf{Medial Hierarchy:} here the second lightest CP-even scalar is the SM-like Higgs.
\end{itemize}
Even within a given hierarchy, the naming of the Higgs states ($H_1,H_2,H_3$) remains a matter of convention. To maintain consistency and simplify the discussion, we will denote the SM-like state in the Regular Hierarchy as $H_1$ and in the Medial Hierarchy as $H_2$. With this framework in place, we now proceed to analyze each hierarchy case in detail, focusing on how the alignment condition constrains the model parameters.
\subsection{Regular Hierarchy}
\label{subsec:reghierarchy}
The coupling of the $H_1$ with the SM gauge bosons in the democratic 3HDM is given by 
\begin{equation}
g_{HZZ}=\frac{ve^2(c_{\beta 2}c_{\alpha 2}c_{(\alpha_1-\beta_1)}+s_{\beta 2}s_{\alpha 2})}{2c_w^2s_w^2}.
\end{equation}
Here we use the shorthand notation $c_\alpha \equiv \cos \alpha$ and $s_\alpha \equiv \sin \alpha$; analogous definitions apply for the other mixing angles. Thus the alignment limit condition for the Regular Hierarchy reads
 \begin{equation}
     	c_{\beta_2}c_{\alpha_2}c_{(\alpha_1-\beta_1)} + s_{\beta_2}s_{\alpha_2} = 1.
      \label{eq:al1}
 \end{equation}
Let $k_1=c_{(\alpha_1-\beta_1)}$ - obviously this parameter can take values in the interval $-1\leq k_1 \leq +1$. However, the case $k_1\neq 1$ corresponds to a distinct configuration in which only the third doublet develops a vev and thus we restrict our analysis to the case $k_1=1$ \cite{Batra:2025amk}. The alignment limit condition thus gives $\alpha_1 = \beta_1 + 2n\pi$ and  $\alpha_2 = \beta_2 + 2n\pi$.
For $k_1=-1$, the scenario is equivalent to the present case, modulo an inversion.

\begin{figure}[h!]
        \includegraphics[scale=0.7]{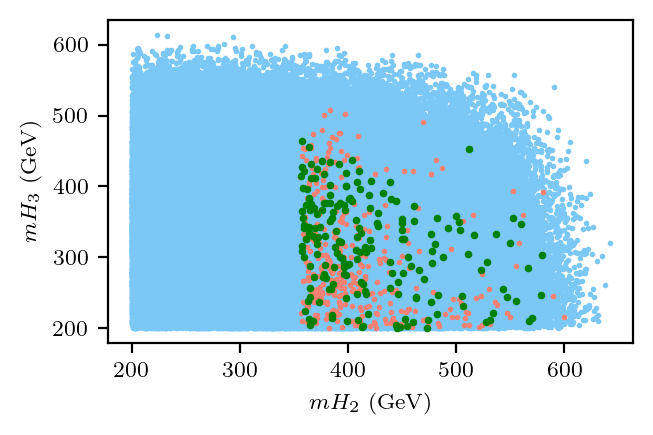}
        \includegraphics[scale=0.7]{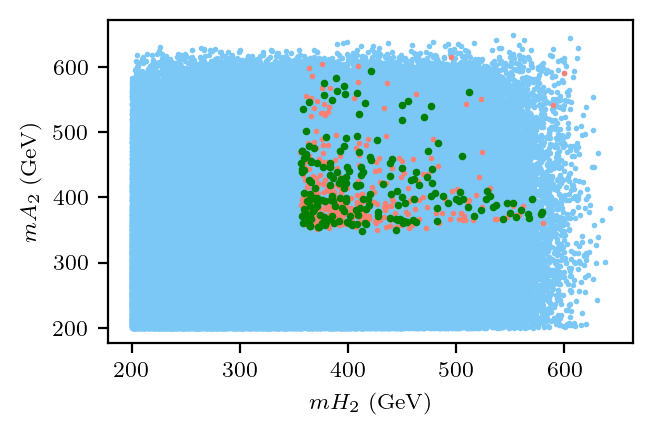}
        \includegraphics[scale=0.7]{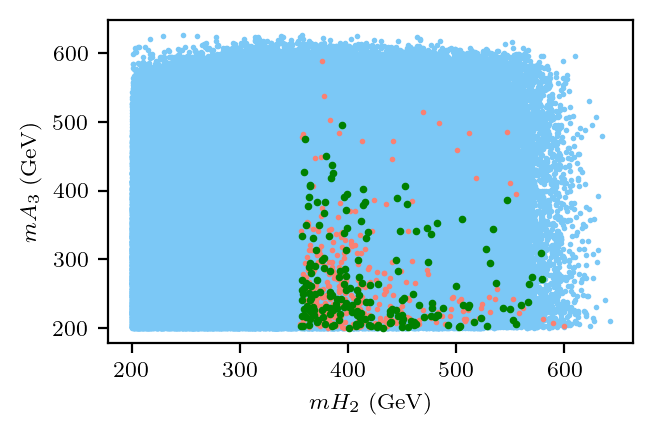}
        \includegraphics[scale=0.7]{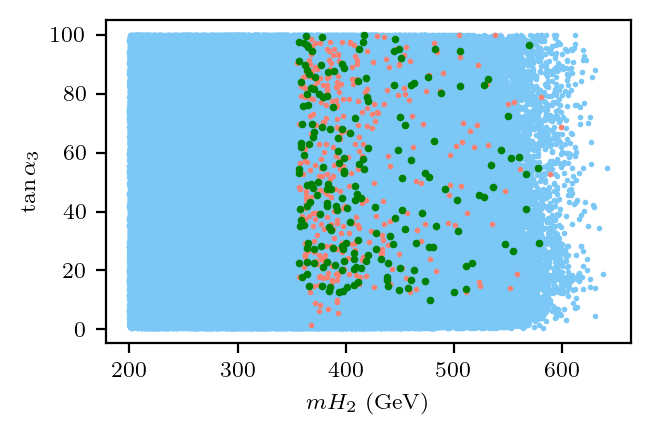}
        \includegraphics[scale=0.7]{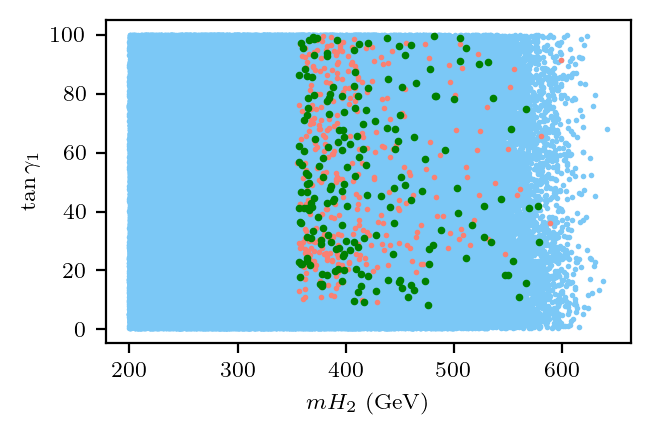}
        \caption{\justifying The allowed regions for the Regular Hierarchy case with the lightest of the three CP-even scalars identified as the 125 GeV SM-like Higgs boson.  The plots show how the allowed region in $m_{H_2}$ correlates with the various angles. The color coding of the plots is as follows: regions shaded in blue is allowed by the stability, perturbativity and unitarity constraints while the red region is allowed by the direct search constraints and the goodness of fit test, and also the $b \rightarrow s \gamma$ flavor constraint. The electroweak precision constraints are indicated by the green shaded region.}
         \label{fig:mH2-al1}
\end{figure}

In Figure~\ref{fig:mH2-al1} (reproduced from \cite{Batra:2025amk}), we present the allowed parameter space for the masses of the additional Higgs states in the model and their correlations with the mixing angles $\gamma_1$, (associated with CP-odd scalar mixing) and $\alpha_3$ (relevant for CP-even Higgs mixing). The color coding of the plots is as follows: regions shaded in blue are allowed by the stability, perturbativity and the unitarity constraints, while the red region is also allowed by the experimental constraints. Regions passing the electroweak precision constraints are marked in green. Thus, the final region marked in green satisfy all the theoretical, experimental, and the electroweak precision constraints simultaneously. From the figure, it is evident that the mass of $H_2$ is tightly constrained to be in the range $350 \ \textrm{GeV} < m_{H_2} < 580 \ \textrm{GeV}$ while $H_3$ lies in the interval $200 \ \textrm{GeV} < m_{H_3} < 460 \ \textrm{GeV}$. The allowed mass ranges for $A_2$ and $A_3$ are found to be comparable. Interestingly, $H_2$ and $A_2$ exhibit similar mass ranges, as do $H_3$ and $A_3$. The mixing angles $\gamma_1$ and $\alpha_3$ on the other hand remain largely unconstrained. 

\subsection{Medial Hierarchy}
In the Medial Hierarchy setup, we identify $H_2$ as the SM-like Higgs boson. This choice imposes a specific mass ordering among the three CP-even Higgs states, placing $H_2$ between a lighter and a heavier scalar. The motivation for considering this scenario is to examine whether the 3HDM framework can accommodate Higgs bosons both lighter and heavier than the SM one, thereby offering interesting collider prospects. To ensure that $H_2$ exhibits SM-like behavior, we require its coupling to the weak gauge bosons to match that of the SM Higgs. Imposing this alignment condition yields the following constraint: 
\begin{equation}
   	c_{\beta_2}c_{\alpha_3}s_{(\beta_1-\alpha_1)} + s_{\beta_2}c_{\alpha_2}s_{\alpha_3} - c_{\beta_2}s_{\alpha_2}s_{\alpha_3}c_{(\alpha_1-\beta_1)} = 1.
\end{equation}
Again letting $k_1=c_{(\alpha_1-\beta_1)}$, this can be re-written as 
\begin{equation}
    \label{eq:medialorder}
   	\pm c_{\beta_2}c_{\alpha_3}\sqrt{1-k_1^2} + s_{\beta_2}c_{\alpha_2}s_{\alpha_3} - c_{\beta_2}s_{\alpha_2}s_{\alpha_3}k_1 = 1.
\end{equation}
Thus the alignment limit condition in this scenario gives, for $k_1=1$, $\alpha_1 = \beta_1 + 2n\pi$.  It is clear from Eqn.~\ref{eq:medialorder} that in this case, we have $\sin\alpha_3\,\sin(\beta_2-\alpha_2)=0$. This has two possible solutions: $\alpha_3 = \frac{\pi}{2}$, $\alpha_2 = \beta_2 - \frac{\pi}{2}$ and $\alpha_3 = \frac{3\pi}{2}$, $\alpha_2 = \beta_2 - \frac{3\pi}{2}$. We can combine these conditions to read $\tan \alpha_2 =  -\cot \beta_2 $. 

\begin{figure}[h!]
        \includegraphics[scale=0.8]{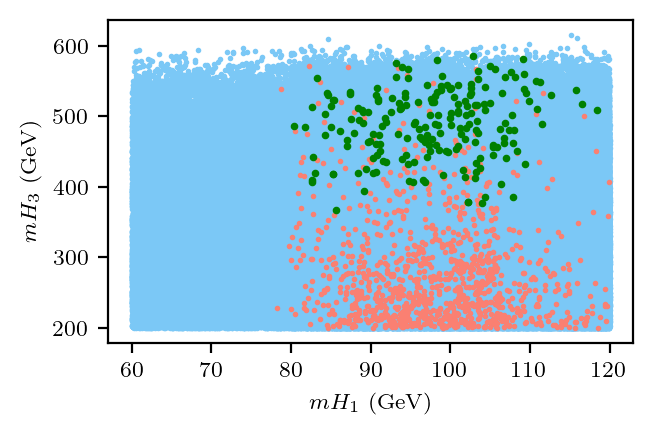}
        \includegraphics[scale=0.8]{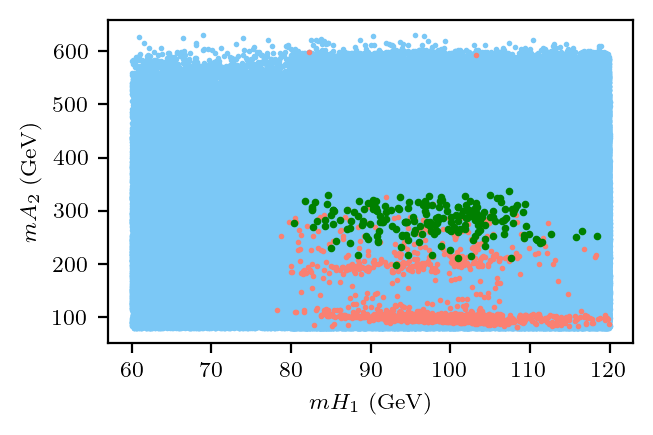}
        \includegraphics[scale=0.8]{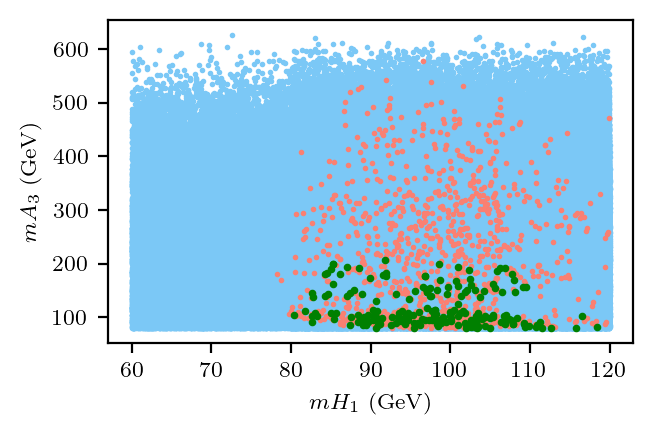}
        \includegraphics[scale=0.8]{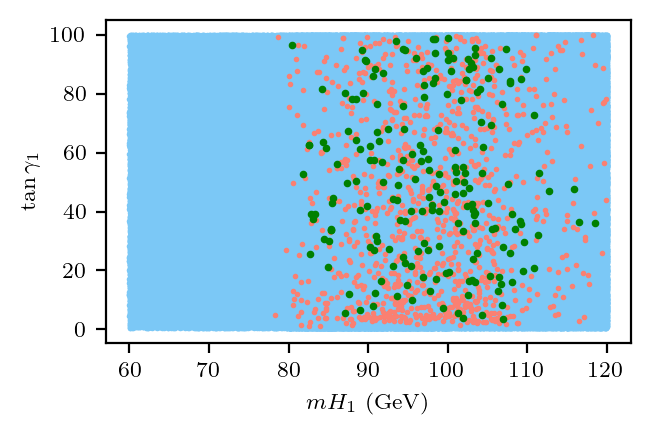}
    
        \caption{\justifying The allowed regions for the Medial Hierarchy case with the second lightest CP-even Higgses identified as the 125 GeV SM-like Higgs boson.  The plots show how the allowed region in $m_{H_1}$ correlates with the masses of the other Higgses in the theory.}
        \label{fig:mH1-al2}
\end{figure}

In Figure~\ref{fig:mH1-al2}, we show the allowed parameter space for the neutral scalar masses together with the mixing angle $\gamma_1$. The color coding follows that of Figure~\ref{fig:mH2-al1}, as described in Section~\ref{subsec:reghierarchy}. For the democratic 3HDM, we find that a light CP-even scalar with $82 \ \textrm{GeV} < m_{H_1} < \textrm{120} \ \textrm{GeV}$ is permitted. The heaviest CP-even Higgs lies in the range $400 \ \textrm{GeV} < m_{H_3} < 600 \ \textrm{GeV}$, comparable to the allowed $H_2$ in the Regular Hierarchy case. In contrast, the CP-odd sector is more tightly constrained, with $200\, \textrm{GeV}\lesssim m_{A_2} \lesssim 320\, \textrm{GeV}$ and $90\, \textrm{GeV}\lesssim m_{A_3} \lesssim 200\, \textrm{GeV}$. The mixing angle $\gamma_1$, however, remains largely unconstrained. Based on the mass hierarchies, the only phenomenologically viable cascade Higgs decay channels of the heavier neutral Higgs states in this scenario are $A_2 \rightarrow ZH_1$ and $A_3 \rightarrow ZH_1$. 
\section{Couplings}
\label{sec:couplings}
In this section, we calculate all the relevant couplings. The couplings of the type $AHZ$ are summarized in Table~\ref{tab:couplings} (where $k_2  =  \sin(\alpha_1-\beta_1)$). Each expression is given up to an overall factor of $(e/2)\csc(\theta_W)\sec(\theta_W)$. For completeness, we also present their simplified forms after imposing the alignment limit conditions. We find that the coupling of the CP-odd Higgs boson with the SM-like Higgs and the $Z$ boson vanishes identically once we impose the alignment limit. On the other hand, its couplings with the remaining two Higgs states exhibit a complementary behavior: if one coupling is enhanced, the other is suppressed. This inverse relation arises solely from the dependence on the mixing angles $\gamma_1$ and $\alpha_3$.
\begin{table}[h]
    \centering
    \begin{tabular}{l l l l}
    \toprule[1pt]
        Coupling & \ \ \ \ \ Coefficient & \ \ \ \ \ Regular & \ \ \ \ \ Medial \\
        \midrule[1pt]
        $A_2H_1Z$ & \ \ \ \ \  $c_{\beta_2}s_{\alpha_2}s_{\gamma_1}-c_{\alpha_2}(c_{\gamma_1}k_2+s_{\beta_2}s_{\gamma_1}k_1)$ & \ \ \ \ \ $0$ & \ \ \ \ \ $-s_{\gamma_1}$ \\
        $A_2H_2Z$ & \ \ \ \ \  $c_{\alpha_3}(-c_{\gamma_1}k_1+s_{\beta_2}s_{\gamma_1}k_2) + s_{\alpha_3}(c_{\gamma_1}s_{\alpha_2}k_2+s_{\gamma_1}(c_{\alpha_2}c_{\beta_2}+s_{\alpha_2}s_{\beta_2}k_1))$ & \ \ \ \ \ $- c_{(\gamma_1+\alpha_3)}$ & \ \ \ \ \ $0$ \\
        $A_2H_3Z$ & \ \ \ \ \  $-s_{\alpha_3}s_{\beta_2}s_{\gamma_1}k_2 + c_{\alpha_3}(s_{\alpha_2}c_{\gamma_1}k_2+c_{\alpha_2}c_{\beta_2}s_{\gamma_1}) + k_1(s_{\alpha_3}c_{\gamma_1}+c_{\alpha_3}s_{\alpha_2}s_{\beta_2}s_{\gamma_1})$ & \ \ \ \ \ $s_{(\gamma_1+\alpha_3)}$ & \ \ \ \ \ $c_{\gamma_1}$ \\
        $A_3H_1Z$ & \ \ \ \ \  $-(c_{\beta_2}s_{\alpha_2}c_{\gamma_1}+c_{\alpha_2}(-s_{\beta_2}c_{\gamma_1}k_1+s_{\gamma_1}k_2))$ & \ \ \ \ \ $0$ & \ \ \ \ \ $c_{\gamma_1}$ \\
        $A_3H_2Z$ & \ \ \ \ \  $-(c_{\alpha_2}c_{\beta_2}c_{\gamma_1}s_{\alpha_3}+k_1(s_{\alpha_2}s_{\beta_2}s_{\alpha_3}c_{\gamma_1}+s_{\gamma_1}c_{\alpha_3}) + k_2(s_{\beta_2}c_{\alpha_3}c_{\gamma_1}-s_{\alpha_2}s_{\alpha_3}s_{\gamma_1}))$ & \ \ \ \ \ $-s_{(\gamma_1+\alpha_3)}$ & \ \ \ \ \ $0$ \\
        $A_3H_3Z$ & \ \ \ \ \  $-c_{\alpha_2}c_{\beta_2}c_{\alpha_3}c_{\gamma_1} + s_{\alpha_3}(s_{\beta_2}c_{\gamma_1}k_2+s_{\gamma_1}k_1)+c_{\alpha_3}s_{\alpha_2}(-s_{\beta_2}c_{\gamma_1}k_1+s_{\gamma_1}k_2)$ & \ \ \ \ \ $- c_{(\gamma_1+\alpha_3)}$ & \ \ \ \ \ $s_{\gamma_1}$ \\
        $A_2t\bar{t}$ & \ \ \ \ \  $c_{\beta_2}s_{\gamma_1}$ & \ \ \ \ \ $c_{\beta_2}s_{\gamma_1}$ & \ \ \ \ \ $c_{\beta_2}s_{\gamma_1}$ \\
        $A_3t\bar{t}$ & \ \ \ \ \  $c_{\beta_2}c_{\gamma_1}$ & \ \ \ \ \ $c_{\beta_2}c_{\gamma_1}$ & \ \ \ \ \ $c_{\beta_2}c_{\gamma_1}$ \\
    \bottomrule[1pt]
    \bottomrule[1pt]
    \end{tabular}
    \caption{\justifying The table gives all couplings of the form $g_{AHZ}$ in the democratic 3HDM that are relevant to the phenomenological analysis in this work. The last two columns give the simplified values of the couplings once the alignment limit corresponding to the Regular or Medial Hierarchies is imposed.}
    \label{tab:couplings}
\end{table}

\section{Collider Phenomenology}
\label{sec:pheno}
\begin{figure}[h!]
    \centering
    \includegraphics[scale=1.0]{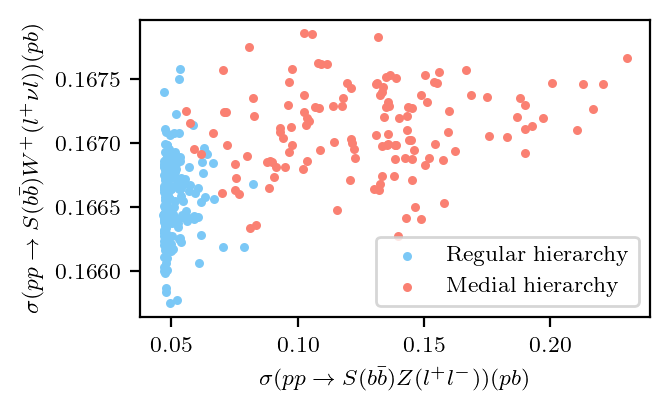}    
    \caption{\justifying The x-axis shows the cross-section for the process $pp \rightarrow S Z \rightarrow b \bar{b} l^+ l^-$, while the y-axis displays the cross-section for the process $pp \rightarrow S W^+ \rightarrow b \bar{b} l^+ \nu$, where $S$ denotes a neutral scalar particle.}
    \label{fig:hierarchyplot}
\end{figure}

In this section, we examine the collider search prospects for the CP-odd and CP-even Higgs bosons within a cascade decay topology. Figure~\ref{fig:hierarchyplot} shows the cross section for the process $pp \rightarrow SV$, where the scalar $S$ subsequently decays into a pair of bottom quarks, and the vector boson $V$ decays leptonically. Here, $S$ collectively denotes all the neutral scalars in the model, including both the CP-even and CP-odd Higgs bosons. The figure includes all benchmark scenarios that satisfy the relevant theoretical and experimental constraints. The cross sections are color-coded: blue represents the Regular Hierarchy, while red corresponds to the Medial Hierarchy. As evident from the plot, the cross sections in the Medial Hierarchy are widely distributed, whereas those in the Regular Hierarchy are clustered within a narrower range along the x-axis. 

\begin{figure}[h!]
    \centering
    \includegraphics[scale=0.5]{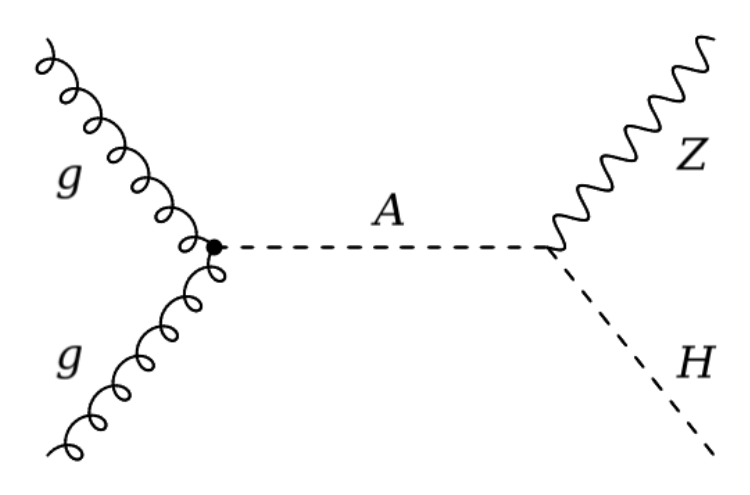}    
    \caption{\justifying Feynman diagram illustrating the gluon fusion production of the CP-odd scalar, with its subsequent decay to a CP-even scalar and a Z boson. The effective coupling of the $A$ to $gg$, denoted by the circle, involves both top and bottom loop contributions. }
    \label{fig:feynman}
\end{figure}

To progress further, we select a representative benchmark point from Fig.~\ref{fig:hierarchyplot} and analyze the collider search prospects for the process $pp\to A \to HZ$, represented in Figure \ref{fig:feynman}, with the $H$ decaying to $b\bar{b}$ and the $Z$ decaying leptonically yielding the final state $b\bar{b}\ell^+\ell^-$ - the Feynman diagram for this process is given in Figure~\ref{fig:feynman}. A detailed phenomenological analysis is carried out for benchmark points that satisfy all relevant theoretical and experimental constraints in both the Regular and the Medial Hierarchy cases. We note here that while the cross-sections for the $pp\to S W^\pm$ process is also comparable, the dominant contribution here is from processes with a $W^\pm$ exchange owing to the presence of the $WWH_1$ (or, $WWH_2$ in the case of Medial Hierarchy) coupling, and thus there is very little change in this cross-section across various benchmark points. The new physics contribution arises with $H_{2,3}^\pm$ in the $s$-channel, but the small cross-sections involved  coupled with the presence of neutrinos in the final state  will make an exact reconstruction of the Higgs, and discovery in this channel, quite challenging and hence we concentrate on the $HZ$ final state in this paper.

\begin{figure}[h!]                     
\includegraphics[scale=0.8]{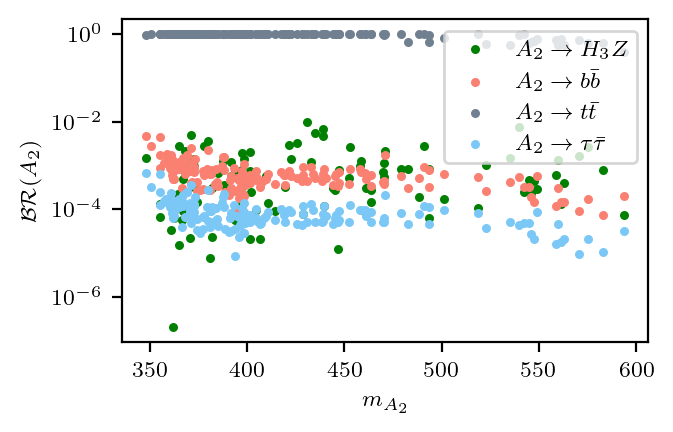}
\includegraphics[scale=0.8]{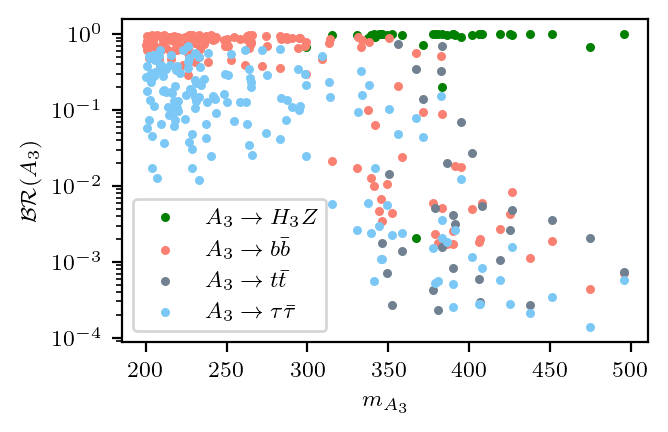}
\includegraphics[scale=0.85]{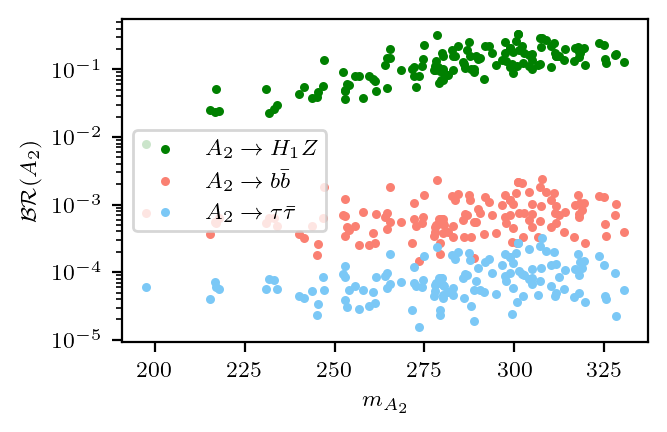}
     \caption{\justifying The branching ratio of the cascade decay of a CP-odd scalar into a CP-even Higgs, plotted as a function of the CP-odd scalar mass. For the Regular Hierarchy (top panel), the branching ratio of both CP-odd scalar is displayed, whereas for the Medial Hierarchy (bottom panel) the corresponding branching ratio of the state $A_2$ is presented.   }
        \label{fig:BR}
\end{figure}

Before selecting specific benchmark points for a detailed collider analysis, it is instructive to examine the decay patterns of the CP-odd Higgs bosons. In Figure~\ref{fig:BR}, we present the branching ratios for the $A \to HZ$ decay channel for all phenomenologically allowed parameter points. For the Regular Hierarchy, we show both viable decay modes, $A_2 \to H_3Z$ as well as $A_3 \to H_3Z$, indicated by blue and red markers, respectively. The figure clearly reveals a complementary behavior between these two channels: when one branching ratio is enhanced, the other is suppressed. This feature can be traced back to the structure of the scalar couplings and is further quantified in Table~\ref{tab:couplings}. In contrast, for the Medial Hierarchy, only a single decay channel $A_3 \to H_1 Z$ has a significant branching ratio. We find that in both scenarios, the branching ratios are clustered within a relatively narrow region. Consequently, selecting a representative benchmark point from each hierarchy is sufficient to capture the generic phenomenological behavior of the model. Guided by this observation, we choose benchmark points corresponding to the largest branching ratios, which in turn maximize the signal cross sections. The selected benchmark points are presented in Table~\ref{tab:benchmarkpoint} and corresponding signal cross sections for these scenarios are presented in Table~\ref{tab:signalxs}.
\begin{table}[h!]
    \centering
    \resizebox{\textwidth}{!}{
    \begin{tabular}{l l l l l l l l l l l l l l l}
    \toprule[1pt]
        BP & \ \ \ $mH_1$ & \ \ \  $mH_2$ & \ \ \ $mH_3$ & \ \ \ $mA_2$ & \ \ \ $mA_3$ & \ \ \ $mH^\pm_2$ & \ \ \ $mH^\pm_3$ & \ \ \ $\tan \beta_1$ & \ \ \ $\tan \beta_2$ & \ \ \ $\tan \alpha_1$ & \ \ \ $\tan \alpha_2$ & \ \ \ $\tan \alpha_3$ & \ \ \ $\tan \gamma_1$ & \ \ \ $\tan \gamma_2$ \\
        \midrule[1pt]
        BP-1 & \ \ \ 125 & \ \ \ 414.07 & \ \ \ 262.19 & \ \ \ 397.33 & \ \ \ 401.90 & \ \ \ 503.57 & \ \ \ 373.77 & \ \ \ 1.06 & \ \ \ 1.04 & \ \ \ 1.06 & \ \ \ 1.04 & \ \ \ 20.71 & \ \ \ 14.88 & \ \ \ 28.18 \\ 
        BP-2 & \ \ \ 85.26 & \ \ \ 125 & \ \ \ 417.83 & \ \ \ 300.94   & \ \ \  199.68 & \ \ \ 363.82 & \ \ \ 378.24 & \ \ \ 1.93 & \ \ \ 2.73 & \ \ \ 1.93 & \ \ \ -0.37 & \ \ \ $\infty$ & \ \ \ 33.65 & \ \ \ 16.48 \\ 
    \bottomrule[1pt]
    \bottomrule[1pt]
    \end{tabular}}
    \caption{\justifying The chosen benchmark points for both the Regular (BP-1) and Medial (BP-2) Hierarchy scenarios discussed in the paper.}
    \label{tab:benchmarkpoint}
\end{table}

\begin{table}[h!]
    \centering    
    \begin{tabular}{l l l}
    \toprule[1pt]
        BP & \ \ \ \ \ \ \ Channel & \ \ \ \ \ \ \ Cross-section(in pb) \\
        \midrule[1pt]
        BP-1 & \ \ \ \ \ \ \ $pp \rightarrow A \rightarrow H_3 Z$ & \ \ \ \ \ \ \ 0.0238 \\
        & \ \ \ \ \ \ \ $pp \rightarrow S \rightarrow S^\prime Z$ & \ \ \ \ \ \ \ 0.0703 \\
        BP-2 & \ \ \ \ \ \ \ $pp \rightarrow A_2 \rightarrow H_1 Z$ & \ \ \ \ \ \ \ 0.09709 \\
        & \ \ \ \ \ \ \ $pp \rightarrow S \rightarrow S^\prime Z$ & \ \ \ \ \ \ \ 0.1435 \\
    \bottomrule[1pt]
    \bottomrule[1pt]
    \end{tabular}
    \caption{\justifying Signal production cross sections for the two benchmark points considered in this study. For BP-1, we analyze a degenerate scenario in which $A$ denotes both the CP-odd neutral scalars, while $S$ and $S^\prime$ represent all the neutral scalar states. In both benchmark points, the produced CP-even scalar subsequently decays into a $b \bar{b}$ pair, and the $Z$ boson decays leptonically. The cross sections quoted correspond to the full inclusive rates, including all decay branching fractions.}
    \label{tab:signalxs}
\end{table}

The event generation for this analysis was carried out using the \texttt{MADGRAPH5\_aMC@NLO} \cite{Alwall:2011uj} event generator. The SM background processes considered in this study are listed in Table~\ref{tab:backchannel}, and were simulated using the default SM implementation provided in the \texttt{MADGRAPH} repository at the $14$ TeV center of mass energy. For the signal events, a model file was constructed using the \texttt{SARAH} \cite{Staub:2008uz} package, and \texttt{SPheno} \cite{Porod:2003um} was employed to provide the spectrum and parameter input. The parton-level events generated by \texttt{MADGRAPH} were subsequently passed to \texttt{PYTHIA} \cite{Bierlich:2022pfr} for parton showering and hadronization. Detector-level effects were simulated using \texttt{DELPHES} \cite{deFavereau:2013fsa}, and the final-state objects were reconstructed with \texttt{MADANALYSIS5} \cite{Conte:2012fm}, which was also used to perform the cut-and-count analysis described in the following section. We carry out the phenomenological analysis for both the benchmark points at an integrated luminosity of $200 \ \textrm{fb}^{-1}$. To initiate the analysis, the following baseline selection of object identification criteria was imposed at the simulation level: 
\begin{equation}
    p_T^j > 20 \; \textrm{GeV}, \; \; \; \; p_T^l > 10 \; \textrm{GeV}, \; \; \; \; |\eta_j| \leq 5, \; \; \; \; |\eta_l| \leq 2.5. 
\end{equation}
 
\begin{table}[h!]
    \centering
    \begin{tabular}{l l l l}
    \toprule[1pt]
        Channel & \ \ \ \ \ \ \ Final State & \ \ \ \ \ \ \ Cross Section (in pb) \\
        \midrule[1pt]
        $pp \rightarrow t \Bar{t} + nj$ & \ \ \ \ \ \ \ $b \Bar{b} \ l^+ l^- \nu l \ \Bar{\nu l}$ & \ \ \ \ \ \ \ $79.5$  \\
        $pp \rightarrow b \Bar{b} + l^+ l^-$ & \ \ \ \ \ \ \ $b \Bar{b} \ l^+ l^-$ & \ \ \ \ \ \ \ $58.5$  \\
        $pp \rightarrow w^+ w^- + nj$ & \ \ \ \ \ \ \ $l^+ l^- \nu l \ \Bar{\nu l}$ & \ \ \ \ \ \ \ $6.13$  \\
        $pp \rightarrow w^\pm z + nj$ & \ \ \ \ \ \ \ $l^\pm \nu l \ b \Bar{b}$ & \ \ \ \ \ \ \ $1.978$  \\
        $pp \rightarrow w^\pm z + nj$ & \ \ \ \ \ \ \ $j j \ l^+ l^-$ & \ \ \ \ \ \ \ $2.73$  \\
        $pp \rightarrow z z + nj$ & \ \ \ \ \ \ \ $l^+ l^- jj$ & \ \ \ \ \ \ \ $1.23$  \\
        $pp \rightarrow t w^\pm + nj$ & \ \ \ \ \ \ \ $l^+ l^- \nu l \ b $ & \ \ \ \ \ \ \ $5.59$  \\
        $pp \rightarrow z + (c \Bar{c}, cc)$ & \ \ \ \ \ \ \ $c \Bar{c}  \ l^+ l^-$ & \ \ \ \ \ \ \ $5.34$  \\
        $pp \rightarrow W^\pm + (b \Bar{b}, c \Bar{c})$ & \ \ \ \ \ \ \ $b \Bar{b} (c \Bar{c}) \ l^\pm \nu_l$ & \ \ \ \ \ \ \ $96.38$  \\
    \bottomrule[1pt]
    \bottomrule[1pt]
    \end{tabular}
    \caption{\justifying The list of all possible background processes (along with their cross-sections) that are relevant to the specific signal process $pp\to A \to HZ\to b\bar{b}\ell^+\ell^- $ considered in this paper.}
    \label{tab:backchannel}
\end{table}
\subsection{Regular Hierarchy}
The cut-flow for the Regular-Hierarchy benchmark point BP-1 is presented in Table~\ref{tab:reghierarchy-sig2-cut1-full} and \ref{tab:reghierarchy-sig2-cut2-full}. This benchmark corresponds to a nearly degenerate CP-odd scalar with a mass of approximately $400~\mathrm{GeV}$, decaying into a $260~\mathrm{GeV}$ CP-even Higgs boson and a $Z$ boson. The resulting signal topology\footnote{We have not included those diagrams with $H \to VV^*$ and $Z \to b \bar{b}$ as the cross-section for this process is significantly lower than the signal cross-section and thus will not affect our analysis.} is characterized by two $b$-jets originating from the Higgs decay and a dilepton pair from the $Z$ boson. We begin by imposing the core object-selection criteria, requiring exactly two $b$-tagged jets $(N_b=2)$ and one positively and one negatively charged lepton $(N_{\ell^+}=N_{\ell^-}=1)$. This selection enforces the expected signal topology and efficiently suppresses a large fraction of the SM backgrounds. A subsequent invariant mass cut on the lepton pair around the $Z$-boson mass window is applied to isolate events containing a resonant dilepton pair, thereby reducing non-resonant and misidentified lepton backgrounds. To further suppress backgrounds involving neutrinos, such as $t\bar{t}$ and $W^+W^-$ production, a missing transverse energy requirement of $E_T^{\rm miss}<20~\mathrm{GeV}$ is imposed.
\begin{figure}[h!]
    \centering
    \includegraphics[scale=0.8]{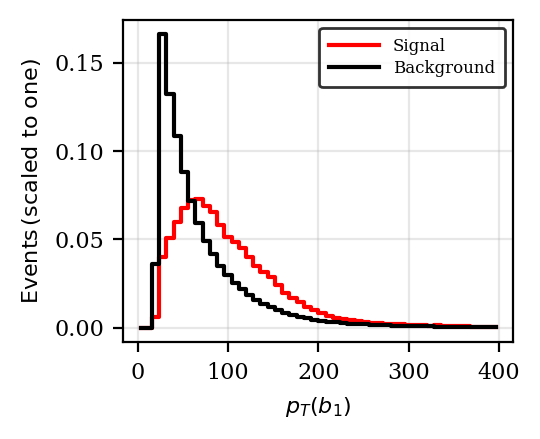}
    \includegraphics[scale=0.8]{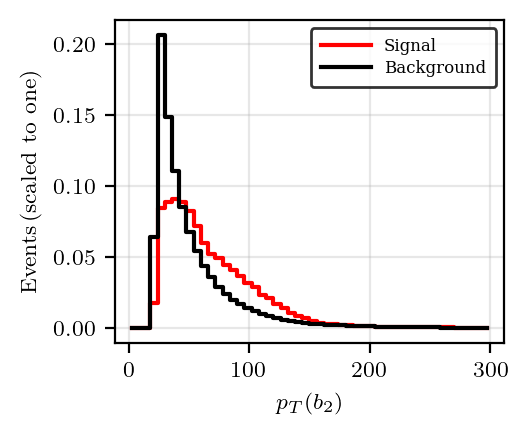}
    \includegraphics[scale=0.8]{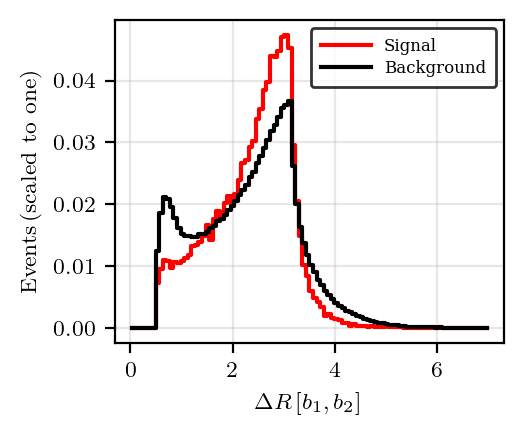}
    \includegraphics[scale=0.8]{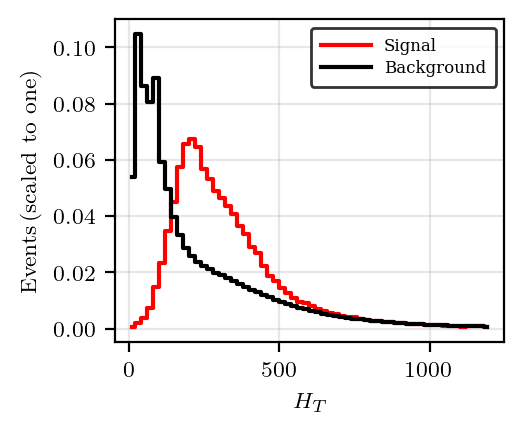}
    \includegraphics[scale=0.8]{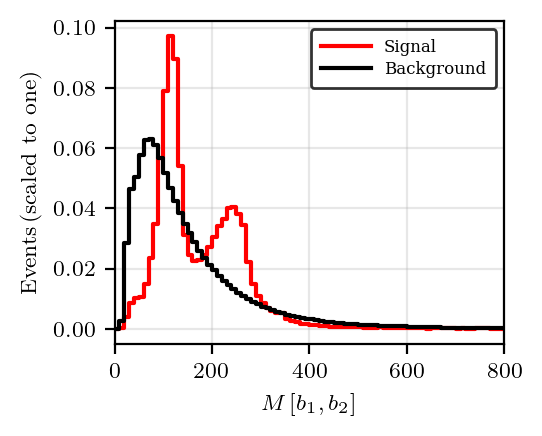}
    \includegraphics[scale=0.8]{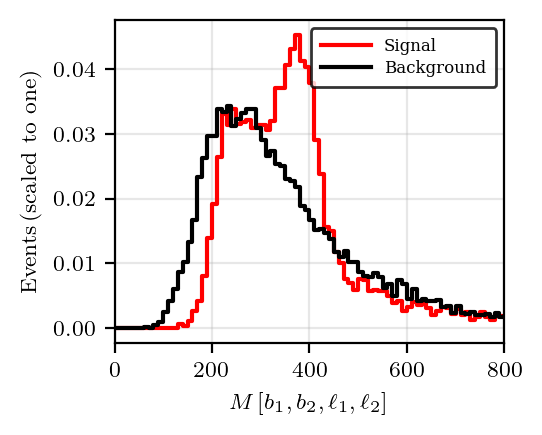}
    \caption{Distributions of the various kinematic observables relevant to the suppression of the SM backgrounds. }
    \label{fig:kinobsreg}
\end{figure}

Following the suppression of neutrino-rich backgrounds through the missing transverse energy cut, we exploit the kinematic features of the heavy scalar decay to further enhance the signal sensitivity. Since the Higgs boson arises from the decay of a relatively heavy CP-odd scalar, the resulting $b$-jets are expected to be moderately boosted. We therefore impose a lower bound on the transverse momentum of the leading $b$-jet, $p_T(b_1) > 60~\mathrm{GeV}$, which significantly reduces residual backgrounds while retaining a large fraction of the signal. Next, we apply a requirement on the angular separation between the two $b$-tagged jets, $0.55 < \Delta R(b_1,b_2) < 3.3$. This cut is motivated by the Higgs decay topology: the lower bound removes collinear configurations often associated with gluon splitting or jet misidentification, while the upper bound suppresses events in which the two $b$-jets are widely separated, a feature more typical of QCD-induced backgrounds. This angular selection further improves the signal-to-background ratio with minimal loss in signal efficiency. Finally, to directly target the resonant production of the heavy CP-odd scalar and the CP-Even scalar, we reconstruct the invariant mass of the full visible final state, presented in Table \ref{tab:reghierarchy-sig2-cut1-full} and \ref{tab:reghierarchy-sig2-cut2-full}. After all selection criteria are applied, a signal-to-background significance of $S/\sqrt{S+B} \simeq 1.58$ is achieved at an integrated luminosity of $200~\mathrm{fb}^{-1}$ for the CP-odd scalar channel and $S/\sqrt{S+B} \simeq 1.50$ for the CP-even scalar channel. Consequently, a $5\sigma$ discovery would require an integrated luminosity of approximately $2200~\mathrm{fb}^{-1}$, which lies well within the projected reach of the HL-LHC program.

\begin{table}[h!]
    \centering
    \resizebox{\textwidth}{!}{
    \begin{tabular}{l l l l l l l l l l l l l}
    \toprule[1pt]
        Cuts & \ \ \ \  Signal & \ \ \ \ $t\bar{t}$ & \ \ \ \   $b\bar{b}l^+l^-$ & \ \ \ \   $w^+w^-$ & \ \ \ \  $wz$ & \ \ \ \  $wz$ & \ \ \ \  $zz+nj$ & \ \ \ \ $tw^\pm$ & \ \ \ \  $z(c\bar{c})$ & \ \ \ \  $w^\pm(b\bar{b)}$ & \ \ \ \ Total & \ \ \ \ S vs B \\
        \midrule[1pt]
        Initial & \ \ \ \ $14065$ & \ \ \ \ $15914479$ & \ \ \ \ $11708859$ & \ \ \ \ $1225520$ & \ \ \ \ $395621$ & \ \ \ \ $546706$ & \ \ \ \ $245480$ & \ \ \ \ $1119257$ & \ \ \ \ $1067029$ & \ \ \ \ $19276719$ & \ \ \ \ $51499676$ & \ \ \ \ $1.96$ \\
        $N_b=2$ & \ \ \ \ $3938$ & \ \ \ \ $4991512$ & \ \ \ \ $410786$ & \ \ \ \ $4193$ & \ \ \ \ $74616$ & \ \ \ \ $6014$ & \ \ \ \ $2299$ & \ \ \ \ $151264$ & \ \ \ \ $24521$ & \ \ \ \ $396694$ & \ \ \ \ $6061901$ & \ \ \ \ $1.59$ \\
        $N_{\ell+}=1$ & \ \ \ \ $2401$ & \ \ \ \ $2838888$ & \ \ \ \ $266001$ & \ \ \ \ $2116$ & \ \ \ \ $23983$ & \ \ \ \ $3069$ & \ \ \ \ $1193$ & \ \ \ \ $92674$ & \ \ \ \ $14179$ & \ \ \ \ $123138$ & \ \ \ \ $3365243$ & \ \ \ \ $1.31$ \\
        $N_{\ell-}=1$ & \ \ \ \ $1503$ & \ \ \ \ $1646377$ & \ \ \ \ $176184$ & \ \ \ \ $1185$ & \ \ \ \ $580$ & \ \ \ \ $1715$ & \ \ \ \ $672$ & \ \ \ \ $56864$ & \ \ \ \ $8549$ & \ \ \ \ $1093$ & \ \ \ \ $1893221$ & \ \ \ \ $1.09$ \\
        $M_Z-20<M_{l^+l^-}<M_Z+20$ & \ \ \ \ $1470$ & \ \ \ \ $1040502$ & \ \ \ \ $160797$ & \ \ \ \ $740$ & \ \ \ \ $573$ & \ \ \ \ $1674$ & \ \ \ \ $661$ & \ \ \ \ $35373$ & \ \ \ \ $8358$ & \ \ \ \ $1072$ & \ \ \ \ $1249752$ & \ \ \ \ $1.31$ \\
        $E_T^{miss}<20$ & \ \ \ \ $748$ & \ \ \ \ $55166$ & \ \ \ \ $104934$ & \ \ \ \ $34$ & \ \ \ \ $49$ & \ \ \ \ $675$ & \ \ \ \ $325$ & \ \ \ \ $1900$ & \ \ \ \ $5160$ & \ \ \ \ $166$ & \ \ \ \ $168410$ & \ \ \ \ $1.82$ \\
        $p_T (b_1) > 60$ & \ \ \ \ $642$ & \ \ \ \ $46894$ & \ \ \ \ $46215$ & \ \ \ \ $20$ & \ \ \ \ $36$ & \ \ \ \ $502$ & \ \ \ \ $212$ & \ \ \ \ $1448$ & \ \ \ \ $3083$ & \ \ \ \ $86$ & \ \ \ \ $98496$ & \ \ \ \ $2.04$ \\
        $0.55 <\Delta R(b_1, b_2) < 3.3$ & \ \ \ \ $603$ & \ \ \ \ $41010$ & \ \ \ \ $35762$ & \ \ \ \ $19$ & \ \ \ \ $34$ & \ \ \ \ $430$ & \ \ \ \ $189$ & \ \ \ \ $1201$ & \ \ \ \ $2423$ & \ \ \ \ $76$ & \ \ \ \ $81146$ & \ \ \ \ $2.11$ \\
        $320 <M(b_1 b_2 \ell_1\ell_2)< 430$ & \ \ \ \ $256$ & \ \ \ \ $13813$ & \ \ \ \ $10816$ & \ \ \ \ $5$ & \ \ \ \ $10$ & \ \ \ \ $123$ & \ \ \ \ $51$ & \ \ \ \ $363$ & \ \ \ \ $733$ & \ \ \ \ $24$ & \ \ \ \ $25937$ & \ \ \ \ $1.58$ \\
    \bottomrule[1pt]
    \bottomrule[1pt]
    \end{tabular} }
    \caption{\justifying Cut flow chart for the process $pp \rightarrow SZ \rightarrow b \bar{b}l^+l^-$ in the Regular Hierarchy case. While the 200 fb$^{-1}$ luminosity can only yield a signal significance of 1.58$\sigma$, this signal should be discoverable at the HL-LHC.}
    \label{tab:reghierarchy-sig2-cut1-full}
\end{table}
\begin{table}[h!]
    \centering
    \resizebox{\textwidth}{!}{
    \begin{tabular}{l l l l l l l l l l l l l}
    \toprule[1pt]
        Cuts & \ \ \ \  Signal & \ \ \ \ $t\bar{t}$ & \ \ \ \   $b\bar{b}l^+l^-$ & \ \ \ \   $w^+w^-$ & \ \ \ \  $wz$ & \ \ \ \  $wz$ & \ \ \ \  $zz+nj$ & \ \ \ \ $tw^\pm$ & \ \ \ \  $z(c\bar{c})$ & \ \ \ \  $w^\pm(b\bar{b)}$ & \ \ \ \ Total & \ \ \ \ S vs B \\
        \midrule[1pt]
        Initial & \ \ \ \ $14065$ & \ \ \ \ $15914479$ & \ \ \ \ $11708859$ & \ \ \ \ $1225520$ & \ \ \ \ $395621$ & \ \ \ \ $546706$ & \ \ \ \ $245480$ & \ \ \ \ $1119257$ & \ \ \ \ $1067029$ & \ \ \ \ $19276719$ & \ \ \ \ $51499676$ & \ \ \ \ $1.96$ \\
        $N_b=2$ & \ \ \ \ $3938$ & \ \ \ \ $4991512$ & \ \ \ \ $410786$ & \ \ \ \ $4193$ & \ \ \ \ $74616$ & \ \ \ \ $6014$ & \ \ \ \ $2299$ & \ \ \ \ $151264$ & \ \ \ \ $24521$ & \ \ \ \ $396694$ & \ \ \ \ $6061901$ & \ \ \ \ $1.59$ \\
        $N_{\ell+}=1$ & \ \ \ \ $2401$ & \ \ \ \ $2838888$ & \ \ \ \ $266001$ & \ \ \ \ $2116$ & \ \ \ \ $23983$ & \ \ \ \ $3069$ & \ \ \ \ $1193$ & \ \ \ \ $92674$ & \ \ \ \ $14179$ & \ \ \ \ $123138$ & \ \ \ \ $3365243$ & \ \ \ \ $1.31$ \\
        $N_{\ell-}=1$ & \ \ \ \ $1503$ & \ \ \ \ $1646377$ & \ \ \ \ $176184$ & \ \ \ \ $1185$ & \ \ \ \ $580$ & \ \ \ \ $1715$ & \ \ \ \ $672$ & \ \ \ \ $56864$ & \ \ \ \ $8549$ & \ \ \ \ $1093$ & \ \ \ \ $1893221$ & \ \ \ \ $1.09$ \\
        $M_Z-20<M_{l^+l^-}<M_Z+20$ & \ \ \ \ $1470$ & \ \ \ \ $1040502$ & \ \ \ \ $160797$ & \ \ \ \ $740$ & \ \ \ \ $573$ & \ \ \ \ $1674$ & \ \ \ \ $661$ & \ \ \ \ $35373$ & \ \ \ \ $8358$ & \ \ \ \ $1072$ & \ \ \ \ $1249752$ & \ \ \ \ $1.31$ \\
        $E_T^{miss}<20$ & \ \ \ \ $748$ & \ \ \ \ $55166$ & \ \ \ \ $104934$ & \ \ \ \ $34$ & \ \ \ \ $49$ & \ \ \ \ $675$ & \ \ \ \ $325$ & \ \ \ \ $1900$ & \ \ \ \ $5160$ & \ \ \ \ $166$ & \ \ \ \ $168410$ & \ \ \ \ $1.82$ \\
        $p_T (b_1) > 60$ & \ \ \ \ $642$ & \ \ \ \ $46894$ & \ \ \ \ $46215$ & \ \ \ \ $20$ & \ \ \ \ $36$ & \ \ \ \ $502$ & \ \ \ \ $212$ & \ \ \ \ $1448$ & \ \ \ \ $3083$ & \ \ \ \ $86$ & \ \ \ \ $98496$ & \ \ \ \ $2.04$ \\
        $0.55 <\Delta R(b_1, b_2) < 3.3$ & \ \ \ \ $603$ & \ \ \ \ $41010$ & \ \ \ \ $35762$ & \ \ \ \ $19$ & \ \ \ \ $34$ & \ \ \ \ $430$ & \ \ \ \ $189$ & \ \ \ \ $1201$ & \ \ \ \ $2423$ & \ \ \ \ $76$ & \ \ \ \ $81146$ & \ \ \ \ $2.11$ \\
        $210 <M(b_1 b_2)< 280$ & \ \ \ \ $155$ & \ \ \ \ $6618$ & \ \ \ \ $3466$ & \ \ \ \ $1$ & \ \ \ \ $0$ & \ \ \ \ $43$ & \ \ \ \ $10$ & \ \ \ \ $175$ & \ \ \ \ $271$ & \ \ \ \ $5$ & \ \ \ \ $10588$ & \ \ \ \ $1.50$ \\
    \bottomrule[1pt]
    \bottomrule[1pt]
    \end{tabular} }
    \caption{\justifying Cut flow chart for the process $pp \rightarrow SZ \rightarrow b \bar{b}l^+l^-$ in the Regular Hierarchy case. While the 200 fb$^{-1}$ luminosity can only yield a signal significance of 1.50$\sigma$, this signal should be discoverable at the HL-LHC.}
    \label{tab:reghierarchy-sig2-cut2-full}
\end{table}

\subsection{Medial Hierarchy}

\begin{figure}[h!]
    \centering
    \includegraphics[scale=0.9]{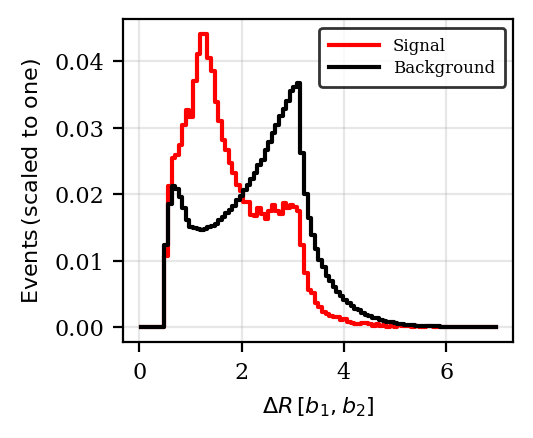}
    \includegraphics[scale=0.9]{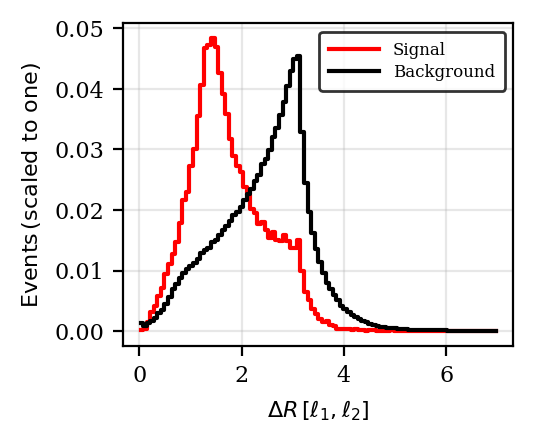}
    \includegraphics[scale=0.9]{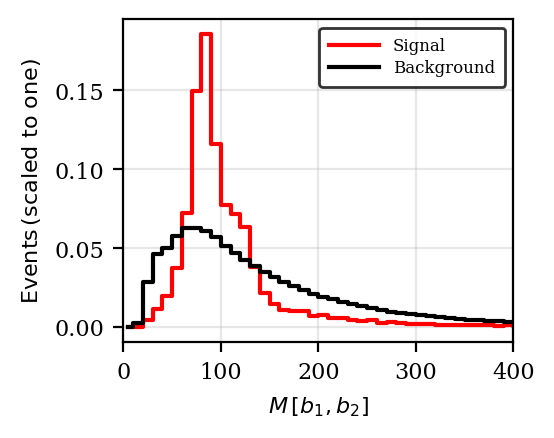}
    \includegraphics[scale=0.9]{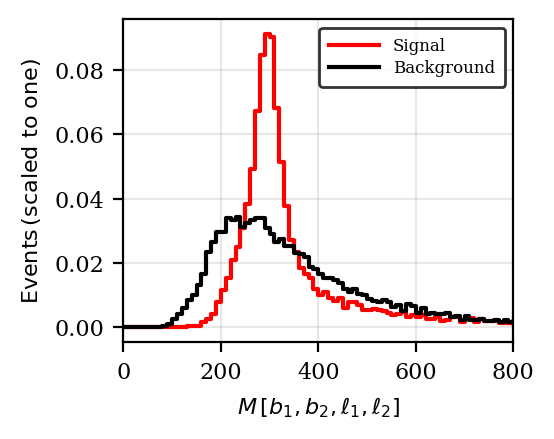}
    \caption{Distribution of various kinematic observables for the Medial Hierarchy}
    \label{fig:kinobsmed}
\end{figure}

The cut-flow for the medium-hierarchy benchmark point, labeled BP-2, is presented in Tables ~\ref{tab:medhierarchy-sig2-cut1-full} and \ref{tab:medhierarchy-sig2-cut2-full}. The selection begins with the basic signal-identification requirements of exactly two $b$-tagged jets and one oppositely charged lepton pair. A tight window around the reconstructed $Z$-boson mass is then applied to suppress backgrounds involving off-shell dileptons or misidentified leptons. Because the signal contains no genuine sources of missing transverse momentum, apart from detector resolution effects, we impose a stringent upper bound of $E_T^{miss} < 20$ GeV. This requirement efficiently removes a large fraction of the $t\bar{t}$ background, for which the neutrinos from $W$ decays typically generate substantial missing energy. A distinctive feature of this process is that both the $H$ and the $Z$ boson originate from the decay of a heavy $300$ GeV pseudoscalar. Consequently, they carry moderate boosts, leading to collimated decay products. We exploit this by applying upper bounds on the angular separations $\Delta R(b_1,b_2)$ and $\Delta R(\ell_1,\ell_2)$. The $\Delta R(b_1,b_2)$ cut suppresses background events in which the two $b$ jets arise from unrelated QCD radiation or from widely separated heavy-flavor production. Likewise, the $\Delta R(\ell_1,\ell_2)$ requirement efficiently rejects dilepton pairs that do not originate from a boosted on-shell $Z$ boson. Finally, we impose mass-window selections on the reconstructed four-body invariant mass $M(b_1 b_2 \ell_1 \ell_2)$, summarized in Table~\ref{tab:medhierarchy-sig2-cut1-full}, and on the di-$b$-jet invariant mass $M(b_1 b_2)$, shown in Table~\ref{tab:medhierarchy-sig2-cut2-full}. These cuts are designed to isolate the $300$ GeV pseudoscalar resonance and the $85$ GeV CP-even Higgs boson, respectively. After applying all selection criteria, the resulting signal yields a discovery significance of approximately $5\sigma$. The kinematic distributions used to motivate and validate these selections are displayed in Fig.~\ref{fig:kinobsmed}.

\begin{table}[h!]
    \centering
    \resizebox{\textwidth}{!}{
    \begin{tabular}{l l l l l l l l l l l l l}
    \toprule[1pt]
            Cuts & \ \ \ \  Signal & \ \ \ \ $t\bar{t}$ & \ \ \ \   $b\bar{b}l^+l^-$ & \ \ \ \   $w^+w^-$ & \ \ \ \  $wz$ & \ \ \ \  $wz$ & \ \ \ \  $zz+nj$ & \ \ \ \ $tw^\pm$ & \ \ \ \  $z(c\bar{c})$ & \ \ \ \  $w^\pm(b\bar{b)}$ & \ \ \ \ Total & \ \ \ \ S vs B \\
        \midrule[1pt]
        Initial & \ \ \ \ $28700$ & \ \ \ \ $15914479$ & \ \ \ \ $11708859$ & \ \ \ \ $1225520$ & \ \ \ \ $395621$ & \ \ \ \ $546706$ & \ \ \ \ $245480$ & \ \ \ \ $1119257$ & \ \ \ \ $1067029$ & \ \ \ \ $19276719$ & \ \ \ \ $51499676$ & \ \ \ \ $3.998$ \\
        $N_b=2$ & \ \ \ \ $6598$ & \ \ \ \ $4991512$ & \ \ \ \ $410786$ & \ \ \ \ $4193$ & \ \ \ \ $74616$ & \ \ \ \ $6014$ & \ \ \ \ $2299$ & \ \ \ \ $151264$ & \ \ \ \ $24521$ & \ \ \ \ $396694$ & \ \ \ \ $6061901$ & \ \ \ \ $2.678$ \\
        $N_{\ell+}=1$ & \ \ \ \ $4370$ & \ \ \ \ $2843929$ & \ \ \ \ $266544$ & \ \ \ \ $2108$ & \ \ \ \ $23464$ & \ \ \ \ $3076$ & \ \ \ \ $1190$ & \ \ \ \ $93040$ & \ \ \ \ $14201$ & \ \ \ \ $121834$ & \ \ \ \ $3369389$ & \ \ \ \ $2.379$ \\
        $N_{\ell-}=1$ & \ \ \ \ $2993$ & \ \ \ \ $1653969$ & \ \ \ \ $176879$ & \ \ \ \ $1176$ & \ \ \ \ $13$ & \ \ \ \ $1725$ & \ \ \ \ $671$ & \ \ \ \ $57261$ & \ \ \ \ $8578$ & \ \ \ \ $43$ & \ \ \ \ $1900316$ & \ \ \ \ $2.169$ \\
        $M_Z-20<M_{l^+l^-}<M_Z+20$ & \ \ \ \ $2920$ & \ \ \ \ $1023899$ & \ \ \ \ $161291$ & \ \ \ \ $724$ & \ \ \ \ $5$ & \ \ \ \ $1682$ & \ \ \ \ $660$ & \ \ \ \ $35120$ & \ \ \ \ $8384$ & \ \ \ \ $22$ & \ \ \ \ $1231787$ & \ \ \ \ $2.628$ \\
        $E_T^{miss}<20$ & \ \ \ \ $1557$ & \ \ \ \ $54707$ & \ \ \ \ $105203$ & \ \ \ \ $34$ & \ \ \ \ $1$ & \ \ \ \ $681$ & \ \ \ \ $328$ & \ \ \ \ $1894$ & \ \ \ \ $5163$ & \ \ \ \ $1$ & \ \ \ \ $168012$ & \ \ \ \ $3.781$ \\
        $0.4<\Delta R(b_1, b_2)< 2.2$ & \ \ \ \ $1179$ & \ \ \ \ $20921$ & \ \ \ \ $35721$ & \ \ \ \ $16$ & \ \ \ \ $1$ & \ \ \ \ $322$ & \ \ \ \ $159$ & \ \ \ \ $752$ & \ \ \ \ $1744$ & \ \ \ \ $0$ & \ \ \ \ $59636$ & \ \ \ \ $4.78$ \\
        $0.4<\Delta R(\ell_1, \ell_2)< 2.2$ & \ \ \ \ $1024$ & \ \ \ \ $12025$ & \ \ \ \ $18437$ & \ \ \ \ $7$ & \ \ \ \ $0$ & \ \ \ \ $218$ & \ \ \ \ $103$ & \ \ \ \ $433$ & \ \ \ \ $1010$ & \ \ \ \ $0$ & \ \ \ \ $32234$ & \ \ \ \ $5.61$ \\
        $270 <M(b_1 b_2 \ell_1\ell_2)< 350$ & \ \ \ \ $670$ & \ \ \ \ $4637$ & \ \ \ \ $5471$ & \ \ \ \ $2$ & \ \ \ \ $0$ & \ \ \ \ $45$ & \ \ \ \ $28$ & \ \ \ \ $128$ & \ \ \ \ $324$ & \ \ \ \ $0$ & \ \ \ \ $10634$ & \ \ \ \ $6.30$ \\
    \bottomrule[1pt]
    \bottomrule[1pt]
    \end{tabular} }
    \caption{\justifying Cut flow chart for the process $pp \rightarrow SZ \rightarrow b \bar{b}l^+l^-$ in the Medial Hierarchy case, illustrating the event selections for reconstructing the CP-odd scalar.}
    \label{tab:medhierarchy-sig2-cut1-full}
\end{table}

\begin{table}[h!]
    \centering
    \resizebox{\textwidth}{!}{
    \begin{tabular}{l l l l l l l l l l l l l}
    \toprule[1pt]
            Cuts & \ \ \ \  Signal & \ \ \ \ $t\bar{t}$ & \ \ \ \   $b\bar{b}l^+l^-$ & \ \ \ \   $w^+w^-$ & \ \ \ \  $wz$ & \ \ \ \  $wz$ & \ \ \ \  $zz+nj$ & \ \ \ \ $tw^\pm$ & \ \ \ \  $z(c\bar{c})$ & \ \ \ \  $w^\pm(b\bar{b)}$ & \ \ \ \ Total & \ \ \ \ S vs B \\
        \midrule[1pt]
        Initial & \ \ \ \ $28700$ & \ \ \ \ $15914479$ & \ \ \ \ $11708859$ & \ \ \ \ $1225520$ & \ \ \ \ $395621$ & \ \ \ \ $546706$ & \ \ \ \ $245480$ & \ \ \ \ $1119257$ & \ \ \ \ $1067029$ & \ \ \ \ $19276719$ & \ \ \ \ $51499676$ & \ \ \ \ $3.998$ \\
        $N_b=2$ & \ \ \ \ $6598$ & \ \ \ \ $4991512$ & \ \ \ \ $410786$ & \ \ \ \ $4193$ & \ \ \ \ $74616$ & \ \ \ \ $6014$ & \ \ \ \ $2299$ & \ \ \ \ $151264$ & \ \ \ \ $24521$ & \ \ \ \ $396694$ & \ \ \ \ $6061901$ & \ \ \ \ $2.678$ \\
        $N_{\ell+}=1$ & \ \ \ \ $4370$ & \ \ \ \ $2843929$ & \ \ \ \ $266544$ & \ \ \ \ $2108$ & \ \ \ \ $23464$ & \ \ \ \ $3076$ & \ \ \ \ $1190$ & \ \ \ \ $93040$ & \ \ \ \ $14201$ & \ \ \ \ $121834$ & \ \ \ \ $3369389$ & \ \ \ \ $2.379$ \\
        $N_{\ell-}=1$ & \ \ \ \ $2993$ & \ \ \ \ $1653969$ & \ \ \ \ $176879$ & \ \ \ \ $1176$ & \ \ \ \ $13$ & \ \ \ \ $1725$ & \ \ \ \ $671$ & \ \ \ \ $57261$ & \ \ \ \ $8578$ & \ \ \ \ $43$ & \ \ \ \ $1900316$ & \ \ \ \ $2.169$ \\
        $M_Z-20<M_{l^+l^-}<M_Z+20$ & \ \ \ \ $2920$ & \ \ \ \ $1023899$ & \ \ \ \ $161291$ & \ \ \ \ $724$ & \ \ \ \ $5$ & \ \ \ \ $1682$ & \ \ \ \ $660$ & \ \ \ \ $35120$ & \ \ \ \ $8384$ & \ \ \ \ $22$ & \ \ \ \ $1231787$ & \ \ \ \ $2.628$ \\
        $E_T^{miss}<20$ & \ \ \ \ $1557$ & \ \ \ \ $54707$ & \ \ \ \ $105203$ & \ \ \ \ $34$ & \ \ \ \ $1$ & \ \ \ \ $681$ & \ \ \ \ $328$ & \ \ \ \ $1894$ & \ \ \ \ $5163$ & \ \ \ \ $1$ & \ \ \ \ $168012$ & \ \ \ \ $3.781$ \\
       $0.4<\Delta R(b_1, b_2)< 2.2$ & \ \ \ \ $1179$ & \ \ \ \ $20921$ & \ \ \ \ $35721$ & \ \ \ \ $16$ & \ \ \ \ $1$ & \ \ \ \ $322$ & \ \ \ \ $159$ & \ \ \ \ $752$ & \ \ \ \ $1744$ & \ \ \ \ $0$ & \ \ \ \ $59636$ & \ \ \ \ $4.78$ \\
        $0.4<\Delta R(\ell_1, \ell_2)< 2.2$ & \ \ \ \ $1024$ & \ \ \ \ $12025$ & \ \ \ \ $18437$ & \ \ \ \ $7$ & \ \ \ \ $0$ & \ \ \ \ $218$ & \ \ \ \ $103$ & \ \ \ \ $433$ & \ \ \ \ $1010$ & \ \ \ \ $0$ & \ \ \ \ $32234$ & \ \ \ \ $5.6$ \\
        $70 <M(b_1 b_2)< 100$ & \ \ \ \ $636$ & \ \ \ \ $3181$ & \ \ \ \ $4584$ & \ \ \ \ $3$ & \ \ \ \ $0$ & \ \ \ \ $72$ & \ \ \ \ $46$ & \ \ \ \ $124$ & \ \ \ \ $252$ & \ \ \ \ $0$ & \ \ \ \ $8263$ & \ \ \ \ $6.74$ \\
    \bottomrule[1pt]
    \bottomrule[1pt]
    \end{tabular} }
    \caption{\justifying Cut flow chart for the process $pp \rightarrow SZ \rightarrow b \bar{b}l^+l^-$ in the Medial Hierarchy case, illustrating the event selections for reconstructing the CP-even scalar.}
    \label{tab:medhierarchy-sig2-cut2-full}
\end{table}

\section{Discussion and Conclusions}
\label{sec:conclusions}

In this work, we have presented a promising collider phenomenology study of the 3HDM with a
$Z_3$-symmetric scalar potential and a Type-Z (democratic) Yukawa structure. Motivated by the rich scalar spectrum
of this framework, we focused on cascade decay signatures involving a CP-odd Higgs boson decaying into a CP-even
scalar and a $Z$ boson, followed by $H \to b\bar{b}$ and leptonic $Z$ decay. Such processes provide a direct probe
of scalar mixing effects and of the extended Higgs structure beyond the SM.

A systematic scan of the parameter space was carried out subject to theoretical constraints from vacuum stability,
perturbative unitarity, and boundedness of the scalar potential, together with experimental bounds from Higgs signal
strengths, direct searches, electroweak precision observables, and flavor physics. Enforcing the alignment limit,
we investigated two phenomenologically viable mass hierarchies: the \emph{Regular Hierarchy}, in which the SM-like
Higgs boson is the lightest CP-even state, and the \emph{Medial Hierarchy}, where the SM-like Higgs occupies an
intermediate position in the spectrum. For each case, representative benchmark points satisfying all constraints
were identified and used for collider analyses. A salient feature of the democratic 3HDM in the alignment limit is that the coupling of the CP-odd Higgs boson to the SM-like Higgs and the $Z$ boson vanishes identically, while its couplings to the remaining CP-even Higgs states
remain nonzero and exhibit a complementary structure. As a result, cascade decays of the form
$pp \to A \to HZ$ naturally emerge as interesting discovery channels for probing the non-SM scalar sector of the
model. These channels are therefore particularly well suited for collider searches targeting extended Higgs
sectors.

We performed a detailed cut-and-count analysis of the $b\bar{b}\ell^+\ell^-$ final state at the LHC with
$\sqrt{s}=14~\text{TeV}$. In the Medial Hierarchy scenario, where a CP-even Higgs lighter than the SM Higgs is
allowed, the resulting kinematic features lead to an efficient background suppression. For the benchmark studied,
a discovery-level statistical significance of approximately $5\sigma$ can be achieved with an integrated
luminosity of $200~\text{fb}^{-1}$, demonstrating strong near-term discovery prospects at the LHC. In contrast, the Regular Hierarchy scenario presents a more challenging experimental environment. Although the same
cascade topology remains viable, the heavier scalar spectrum and reduced kinematic separation from Standard Model
backgrounds significantly suppress the signal yield. For the benchmark considered, we find that an integrated
luminosity of order $2~\text{ab}^{-1}$ is required to reach a $5\sigma$ significance, placing this scenario within
the discovery reach of the High-Luminosity LHC.

In summary, our analysis demonstrates that cascade decays involving pseudoscalar Higgs bosons constitute a powerful
and complementary probe of the scalar sector in the Three Higgs Doublet Model. The strong dependence of the collider
sensitivity on the underlying Higgs mass hierarchy highlights the importance of considering nonstandard Higgs
orderings, particularly scenarios allowing lighter-than-SM Higgs states. A more exhaustive exploration of the 3HDM parameter space (with different Yukawa structures than the one considered in the present study), including alternative Higgs mass hierarchies and complementary cascade decay channels will further elucidate the collider phenomenology and experimental
testability of this framework at the LHC and its future upgrades.

\begin{acknowledgments}
A.K. acknowledges the support from the Director's Fellowship at IIT Gandhinagar. A.S. thanks Anusandhan National Research Foundation (ANRF) for providing necessary financial support through the SERB-NPDF grant (Ref No: PDF/2023/002572). S.K.R. acknowledges the support from the Department of Atomic Energy (DAE), India, for the Regional Centre for Accelerator-based Particle Physics (RECAPP), Harish Chandra Research Institute. 
\end{acknowledgments}

\label{sec:conclusions}

\appendix

\bibliography{bibliography}

\begin{thebibliography}{48}%
\makeatletter
\providecommand \@ifxundefined [1]{%
 \@ifx{#1\undefined}
}%
\providecommand \@ifnum [1]{%
 \ifnum #1\expandafter \@firstoftwo
 \else \expandafter \@secondoftwo
 \fi
}%
\providecommand \@ifx [1]{%
 \ifx #1\expandafter \@firstoftwo
 \else \expandafter \@secondoftwo
 \fi
}%
\providecommand \natexlab [1]{#1}%
\providecommand \enquote  [1]{``#1''}%
\providecommand \bibnamefont  [1]{#1}%
\providecommand \bibfnamefont [1]{#1}%
\providecommand \citenamefont [1]{#1}%
\providecommand \href@noop [0]{\@secondoftwo}%
\providecommand \href [0]{\begingroup \@sanitize@url \@href}%
\providecommand \@href[1]{\@@startlink{#1}\@@href}%
\providecommand \@@href[1]{\endgroup#1\@@endlink}%
\providecommand \@sanitize@url [0]{\catcode `\\12\catcode `\$12\catcode
  `\&12\catcode `\#12\catcode `\^12\catcode `\_12\catcode `\%12\relax}%
\providecommand \@@startlink[1]{}%
\providecommand \@@endlink[0]{}%
\providecommand \url  [0]{\begingroup\@sanitize@url \@url }%
\providecommand \@url [1]{\endgroup\@href {#1}{\urlprefix }}%
\providecommand \urlprefix  [0]{URL }%
\providecommand \Eprint [0]{\href }%
\providecommand \doibase [0]{https://doi.org/}%
\providecommand \selectlanguage [0]{\@gobble}%
\providecommand \bibinfo  [0]{\@secondoftwo}%
\providecommand \bibfield  [0]{\@secondoftwo}%
\providecommand \translation [1]{[#1]}%
\providecommand \BibitemOpen [0]{}%
\providecommand \bibitemStop [0]{}%
\providecommand \bibitemNoStop [0]{.\EOS\space}%
\providecommand \EOS [0]{\spacefactor3000\relax}%
\providecommand \BibitemShut  [1]{\csname bibitem#1\endcsname}%
\let\auto@bib@innerbib\@empty
\bibitem [{\citenamefont {Salam}(1968)}]{Salam:1968rm}%
  \BibitemOpen
  \bibfield  {author} {\bibinfo {author} {\bibfnamefont {A.}~\bibnamefont
  {Salam}},\ }\bibfield  {title} {\bibinfo {title} {{Weak and Electromagnetic
  Interactions}},\ }\href {https://doi.org/10.1142/9789812795915_0034}
  {\bibfield  {journal} {\bibinfo  {journal} {Conf. Proc. C}\ }\textbf
  {\bibinfo {volume} {680519}},\ \bibinfo {pages} {367} (\bibinfo {year}
  {1968})}\BibitemShut {NoStop}%
\bibitem [{\citenamefont {Weinberg}(1967)}]{Weinberg:1967tq}%
  \BibitemOpen
  \bibfield  {author} {\bibinfo {author} {\bibfnamefont {S.}~\bibnamefont
  {Weinberg}},\ }\bibfield  {title} {\bibinfo {title} {{A Model of Leptons}},\
  }\href {https://doi.org/10.1103/PhysRevLett.19.1264} {\bibfield  {journal}
  {\bibinfo  {journal} {Phys. Rev. Lett.}\ }\textbf {\bibinfo {volume} {19}},\
  \bibinfo {pages} {1264} (\bibinfo {year} {1967})}\BibitemShut {NoStop}%
\bibitem [{\citenamefont {Glashow}(1961)}]{Glashow:1961tr}%
  \BibitemOpen
  \bibfield  {author} {\bibinfo {author} {\bibfnamefont {S.~L.}\ \bibnamefont
  {Glashow}},\ }\bibfield  {title} {\bibinfo {title} {{Partial Symmetries of
  Weak Interactions}},\ }\href {https://doi.org/10.1016/0029-5582(61)90469-2}
  {\bibfield  {journal} {\bibinfo  {journal} {Nucl. Phys.}\ }\textbf {\bibinfo
  {volume} {22}},\ \bibinfo {pages} {579} (\bibinfo {year} {1961})}\BibitemShut
  {NoStop}%
\bibitem [{\citenamefont {Akeroyd}\ \emph
  {et~al.}(2021{\natexlab{a}})\citenamefont {Akeroyd}, \citenamefont {Moretti},
  \citenamefont {Shindou},\ and\ \citenamefont {Song}}]{Akeroyd_2021}%
  \BibitemOpen
  \bibfield  {author} {\bibinfo {author} {\bibfnamefont {A.}~\bibnamefont
  {Akeroyd}}, \bibinfo {author} {\bibfnamefont {S.}~\bibnamefont {Moretti}},
  \bibinfo {author} {\bibfnamefont {T.}~\bibnamefont {Shindou}},\ and\ \bibinfo
  {author} {\bibfnamefont {M.}~\bibnamefont {Song}},\ }\bibfield  {title}
  {\bibinfo {title} {Cp asymmetries of $\bar{B} \rightarrow x_s/x_d \gamma$ in
  models with three higgs doublets},\ }\bibfield  {journal} {\bibinfo
  {journal} {Physical Review D}\ }\textbf {\bibinfo {volume} {103}},\ \href
  {https://doi.org/10.1103/physrevd.103.015035} {10.1103/physrevd.103.015035}
  (\bibinfo {year} {2021}{\natexlab{a}})\BibitemShut {NoStop}%
\bibitem [{\citenamefont {Chatrchyan}\ \emph {et~al.}(2012)\citenamefont
  {Chatrchyan} \emph {et~al.}}]{CMS:2012qbp}%
  \BibitemOpen
  \bibfield  {author} {\bibinfo {author} {\bibfnamefont {S.}~\bibnamefont
  {Chatrchyan}} \emph {et~al.} (\bibinfo {collaboration} {CMS}),\ }\bibfield
  {title} {\bibinfo {title} {{Observation of a New Boson at a Mass of 125 GeV
  with the CMS Experiment at the LHC}},\ }\href
  {https://doi.org/10.1016/j.physletb.2012.08.021} {\bibfield  {journal}
  {\bibinfo  {journal} {Phys. Lett. B}\ }\textbf {\bibinfo {volume} {716}},\
  \bibinfo {pages} {30} (\bibinfo {year} {2012})},\ \Eprint
  {https://arxiv.org/abs/1207.7235} {arXiv:1207.7235 [hep-ex]} \BibitemShut
  {NoStop}%
\bibitem [{\citenamefont {Branco}\ \emph {et~al.}(2012)\citenamefont {Branco},
  \citenamefont {Ferreira}, \citenamefont {Lavoura}, \citenamefont {Rebelo},
  \citenamefont {Sher},\ and\ \citenamefont {Silva}}]{Branco:2011iw}%
  \BibitemOpen
  \bibfield  {author} {\bibinfo {author} {\bibfnamefont {G.~C.}\ \bibnamefont
  {Branco}}, \bibinfo {author} {\bibfnamefont {P.~M.}\ \bibnamefont
  {Ferreira}}, \bibinfo {author} {\bibfnamefont {L.}~\bibnamefont {Lavoura}},
  \bibinfo {author} {\bibfnamefont {M.~N.}\ \bibnamefont {Rebelo}}, \bibinfo
  {author} {\bibfnamefont {M.}~\bibnamefont {Sher}},\ and\ \bibinfo {author}
  {\bibfnamefont {J.~P.}\ \bibnamefont {Silva}},\ }\bibfield  {title} {\bibinfo
  {title} {{Theory and phenomenology of two-Higgs-doublet models}},\ }\href
  {https://doi.org/10.1016/j.physrep.2012.02.002} {\bibfield  {journal}
  {\bibinfo  {journal} {Phys. Rept.}\ }\textbf {\bibinfo {volume} {516}},\
  \bibinfo {pages} {1} (\bibinfo {year} {2012})},\ \Eprint
  {https://arxiv.org/abs/1106.0034} {arXiv:1106.0034 [hep-ph]} \BibitemShut
  {NoStop}%
\bibitem [{\citenamefont {Coleppa}\ \emph {et~al.}(2014)\citenamefont
  {Coleppa}, \citenamefont {Kling},\ and\ \citenamefont
  {Su}}]{Coleppa:2013dya}%
  \BibitemOpen
  \bibfield  {author} {\bibinfo {author} {\bibfnamefont {B.}~\bibnamefont
  {Coleppa}}, \bibinfo {author} {\bibfnamefont {F.}~\bibnamefont {Kling}},\
  and\ \bibinfo {author} {\bibfnamefont {S.}~\bibnamefont {Su}},\ }\bibfield
  {title} {\bibinfo {title} {{Constraining Type II 2HDM in Light of LHC Higgs
  Searches}},\ }\href {https://doi.org/10.1007/JHEP01(2014)161} {\bibfield
  {journal} {\bibinfo  {journal} {JHEP}\ }\textbf {\bibinfo {volume} {01}},\
  \bibinfo {pages} {161}},\ \Eprint {https://arxiv.org/abs/1305.0002}
  {arXiv:1305.0002 [hep-ph]} \BibitemShut {NoStop}%
\bibitem [{\citenamefont {Chang}\ \emph {et~al.}(2013)\citenamefont {Chang},
  \citenamefont {Kang}, \citenamefont {Lee}, \citenamefont {Lee}, \citenamefont
  {Park},\ and\ \citenamefont {Song}}]{Chang:2012ve}%
  \BibitemOpen
  \bibfield  {author} {\bibinfo {author} {\bibfnamefont {S.}~\bibnamefont
  {Chang}}, \bibinfo {author} {\bibfnamefont {S.~K.}\ \bibnamefont {Kang}},
  \bibinfo {author} {\bibfnamefont {J.-P.}\ \bibnamefont {Lee}}, \bibinfo
  {author} {\bibfnamefont {K.~Y.}\ \bibnamefont {Lee}}, \bibinfo {author}
  {\bibfnamefont {S.~C.}\ \bibnamefont {Park}},\ and\ \bibinfo {author}
  {\bibfnamefont {J.}~\bibnamefont {Song}},\ }\bibfield  {title} {\bibinfo
  {title} {{Comprehensive study of two Higgs doublet model in light of the new
  boson with mass around 125 GeV}},\ }\href
  {https://doi.org/10.1007/JHEP05(2013)075} {\bibfield  {journal} {\bibinfo
  {journal} {JHEP}\ }\textbf {\bibinfo {volume} {05}},\ \bibinfo {pages}
  {075}},\ \Eprint {https://arxiv.org/abs/1210.3439} {arXiv:1210.3439 [hep-ph]}
  \BibitemShut {NoStop}%
\bibitem [{\citenamefont {Grinstein}\ and\ \citenamefont
  {Uttayarat}(2013)}]{Grinstein:2013npa}%
  \BibitemOpen
  \bibfield  {author} {\bibinfo {author} {\bibfnamefont {B.}~\bibnamefont
  {Grinstein}}\ and\ \bibinfo {author} {\bibfnamefont {P.}~\bibnamefont
  {Uttayarat}},\ }\bibfield  {title} {\bibinfo {title} {{Carving Out Parameter
  Space in Type-II Two Higgs Doublets Model}},\ }\href
  {https://doi.org/10.1007/JHEP06(2013)094} {\bibfield  {journal} {\bibinfo
  {journal} {JHEP}\ }\textbf {\bibinfo {volume} {06}},\ \bibinfo {pages}
  {094}},\ \bibinfo {note} {[Erratum: JHEP 09, 110 (2013)]},\ \Eprint
  {https://arxiv.org/abs/1304.0028} {arXiv:1304.0028 [hep-ph]} \BibitemShut
  {NoStop}%
\bibitem [{\citenamefont {Drozd}\ \emph {et~al.}(2014)\citenamefont {Drozd},
  \citenamefont {Grzadkowski}, \citenamefont {Gunion},\ and\ \citenamefont
  {Jiang}}]{Drozd:2014yla}%
  \BibitemOpen
  \bibfield  {author} {\bibinfo {author} {\bibfnamefont {A.}~\bibnamefont
  {Drozd}}, \bibinfo {author} {\bibfnamefont {B.}~\bibnamefont {Grzadkowski}},
  \bibinfo {author} {\bibfnamefont {J.~F.}\ \bibnamefont {Gunion}},\ and\
  \bibinfo {author} {\bibfnamefont {Y.}~\bibnamefont {Jiang}},\ }\bibfield
  {title} {\bibinfo {title} {{Extending two-Higgs-doublet models by a singlet
  scalar field - the Case for Dark Matter}},\ }\href
  {https://doi.org/10.1007/JHEP11(2014)105} {\bibfield  {journal} {\bibinfo
  {journal} {JHEP}\ }\textbf {\bibinfo {volume} {11}},\ \bibinfo {pages}
  {105}},\ \Eprint {https://arxiv.org/abs/1408.2106} {arXiv:1408.2106 [hep-ph]}
  \BibitemShut {NoStop}%
\bibitem [{\citenamefont {Bhattacharya}\ \emph {et~al.}(2024)\citenamefont
  {Bhattacharya}, \citenamefont {Dey}, \citenamefont {Lahiri},\ and\
  \citenamefont {Mukhopadhyaya}}]{Bhattacharya:2023qfs}%
  \BibitemOpen
  \bibfield  {author} {\bibinfo {author} {\bibfnamefont {S.}~\bibnamefont
  {Bhattacharya}}, \bibinfo {author} {\bibfnamefont {A.}~\bibnamefont {Dey}},
  \bibinfo {author} {\bibfnamefont {J.}~\bibnamefont {Lahiri}},\ and\ \bibinfo
  {author} {\bibfnamefont {B.}~\bibnamefont {Mukhopadhyaya}},\ }\bibfield
  {title} {\bibinfo {title} {{High scale validity of two-Higgs-doublet
  scenarios with a real scalar singlet dark matter}},\ }\href
  {https://doi.org/10.1103/PhysRevD.110.055034} {\bibfield  {journal} {\bibinfo
   {journal} {Phys. Rev. D}\ }\textbf {\bibinfo {volume} {110}},\ \bibinfo
  {pages} {055034} (\bibinfo {year} {2024})},\ \Eprint
  {https://arxiv.org/abs/2308.12473} {arXiv:2308.12473 [hep-ph]} \BibitemShut
  {NoStop}%
\bibitem [{\citenamefont {Muhlleitner}\ \emph {et~al.}(2017)\citenamefont
  {Muhlleitner}, \citenamefont {Sampaio}, \citenamefont {Santos},\ and\
  \citenamefont {Wittbrodt}}]{Muhlleitner:2016mzt}%
  \BibitemOpen
  \bibfield  {author} {\bibinfo {author} {\bibfnamefont {M.}~\bibnamefont
  {Muhlleitner}}, \bibinfo {author} {\bibfnamefont {M.~O.~P.}\ \bibnamefont
  {Sampaio}}, \bibinfo {author} {\bibfnamefont {R.}~\bibnamefont {Santos}},\
  and\ \bibinfo {author} {\bibfnamefont {J.}~\bibnamefont {Wittbrodt}},\
  }\bibfield  {title} {\bibinfo {title} {{The N2HDM under Theoretical and
  Experimental Scrutiny}},\ }\href {https://doi.org/10.1007/JHEP03(2017)094}
  {\bibfield  {journal} {\bibinfo  {journal} {JHEP}\ }\textbf {\bibinfo
  {volume} {03}},\ \bibinfo {pages} {094}},\ \Eprint
  {https://arxiv.org/abs/1612.01309} {arXiv:1612.01309 [hep-ph]} \BibitemShut
  {NoStop}%
\bibitem [{\citenamefont {Keus}\ \emph {et~al.}(2018)\citenamefont {Keus},
  \citenamefont {Koivunen},\ and\ \citenamefont {Tuominen}}]{Keus:2017ioh}%
  \BibitemOpen
  \bibfield  {author} {\bibinfo {author} {\bibfnamefont {V.}~\bibnamefont
  {Keus}}, \bibinfo {author} {\bibfnamefont {N.}~\bibnamefont {Koivunen}},\
  and\ \bibinfo {author} {\bibfnamefont {K.}~\bibnamefont {Tuominen}},\
  }\bibfield  {title} {\bibinfo {title} {{Singlet scalar and 2HDM extensions of
  the Standard Model: CP-violation and constraints from $(g-2)_\mu$ and
  $e$EDM}},\ }\href {https://doi.org/10.1007/JHEP09(2018)059} {\bibfield
  {journal} {\bibinfo  {journal} {JHEP}\ }\textbf {\bibinfo {volume} {09}},\
  \bibinfo {pages} {059}},\ \Eprint {https://arxiv.org/abs/1712.09613}
  {arXiv:1712.09613 [hep-ph]} \BibitemShut {NoStop}%
\bibitem [{\citenamefont {Bhattacharyya}\ and\ \citenamefont
  {Das}(2016)}]{Bhattacharyya:2015nca}%
  \BibitemOpen
  \bibfield  {author} {\bibinfo {author} {\bibfnamefont {G.}~\bibnamefont
  {Bhattacharyya}}\ and\ \bibinfo {author} {\bibfnamefont {D.}~\bibnamefont
  {Das}},\ }\bibfield  {title} {\bibinfo {title} {{Scalar sector of
  two-Higgs-doublet models: A minireview}},\ }\href
  {https://doi.org/10.1007/s12043-016-1252-4} {\bibfield  {journal} {\bibinfo
  {journal} {Pramana}\ }\textbf {\bibinfo {volume} {87}},\ \bibinfo {pages}
  {40} (\bibinfo {year} {2016})},\ \Eprint {https://arxiv.org/abs/1507.06424}
  {arXiv:1507.06424 [hep-ph]} \BibitemShut {NoStop}%
\bibitem [{\citenamefont {Cao}\ \emph {et~al.}(2009)\citenamefont {Cao},
  \citenamefont {Wan}, \citenamefont {Wu},\ and\ \citenamefont
  {Yang}}]{Cao:2009as}%
  \BibitemOpen
  \bibfield  {author} {\bibinfo {author} {\bibfnamefont {J.}~\bibnamefont
  {Cao}}, \bibinfo {author} {\bibfnamefont {P.}~\bibnamefont {Wan}}, \bibinfo
  {author} {\bibfnamefont {L.}~\bibnamefont {Wu}},\ and\ \bibinfo {author}
  {\bibfnamefont {J.~M.}\ \bibnamefont {Yang}},\ }\bibfield  {title} {\bibinfo
  {title} {{Lepton-Specific Two-Higgs Doublet Model: Experimental Constraints
  and Implication on Higgs Phenomenology}},\ }\href
  {https://doi.org/10.1103/PhysRevD.80.071701} {\bibfield  {journal} {\bibinfo
  {journal} {Phys. Rev. D}\ }\textbf {\bibinfo {volume} {80}},\ \bibinfo
  {pages} {071701} (\bibinfo {year} {2009})},\ \Eprint
  {https://arxiv.org/abs/0909.5148} {arXiv:0909.5148 [hep-ph]} \BibitemShut
  {NoStop}%
\bibitem [{\citenamefont {Ko}\ \emph {et~al.}(2014)\citenamefont {Ko},
  \citenamefont {Omura},\ and\ \citenamefont {Yu}}]{Ko:2013zsa}%
  \BibitemOpen
  \bibfield  {author} {\bibinfo {author} {\bibfnamefont {P.}~\bibnamefont
  {Ko}}, \bibinfo {author} {\bibfnamefont {Y.}~\bibnamefont {Omura}},\ and\
  \bibinfo {author} {\bibfnamefont {C.}~\bibnamefont {Yu}},\ }\bibfield
  {title} {\bibinfo {title} {{Higgs phenomenology in Type-I 2HDM with $U(1)_H$
  Higgs gauge symmetry}},\ }\href {https://doi.org/10.1007/JHEP01(2014)016}
  {\bibfield  {journal} {\bibinfo  {journal} {JHEP}\ }\textbf {\bibinfo
  {volume} {01}},\ \bibinfo {pages} {016}},\ \Eprint
  {https://arxiv.org/abs/1309.7156} {arXiv:1309.7156 [hep-ph]} \BibitemShut
  {NoStop}%
\bibitem [{\citenamefont {Crivellin}\ \emph {et~al.}(2013)\citenamefont
  {Crivellin}, \citenamefont {Kokulu},\ and\ \citenamefont
  {Greub}}]{Crivellin:2013wna}%
  \BibitemOpen
  \bibfield  {author} {\bibinfo {author} {\bibfnamefont {A.}~\bibnamefont
  {Crivellin}}, \bibinfo {author} {\bibfnamefont {A.}~\bibnamefont {Kokulu}},\
  and\ \bibinfo {author} {\bibfnamefont {C.}~\bibnamefont {Greub}},\ }\bibfield
   {title} {\bibinfo {title} {{Flavor-phenomenology of two-Higgs-doublet models
  with generic Yukawa structure}},\ }\href
  {https://doi.org/10.1103/PhysRevD.87.094031} {\bibfield  {journal} {\bibinfo
  {journal} {Phys. Rev. D}\ }\textbf {\bibinfo {volume} {87}},\ \bibinfo
  {pages} {094031} (\bibinfo {year} {2013})},\ \Eprint
  {https://arxiv.org/abs/1303.5877} {arXiv:1303.5877 [hep-ph]} \BibitemShut
  {NoStop}%
\bibitem [{\citenamefont {Heinemeyer}\ \emph {et~al.}(2022)\citenamefont
  {Heinemeyer}, \citenamefont {Li}, \citenamefont {Lika}, \citenamefont
  {Moortgat-Pick},\ and\ \citenamefont {Paasch}}]{PhysRevD.106.075003}%
  \BibitemOpen
  \bibfield  {author} {\bibinfo {author} {\bibfnamefont {S.}~\bibnamefont
  {Heinemeyer}}, \bibinfo {author} {\bibfnamefont {C.}~\bibnamefont {Li}},
  \bibinfo {author} {\bibfnamefont {F.}~\bibnamefont {Lika}}, \bibinfo {author}
  {\bibfnamefont {G.}~\bibnamefont {Moortgat-Pick}},\ and\ \bibinfo {author}
  {\bibfnamefont {S.}~\bibnamefont {Paasch}},\ }\bibfield  {title} {\bibinfo
  {title} {Phenomenology of a 96 gev higgs boson in the 2hdm with an additional
  singlet},\ }\href {https://doi.org/10.1103/PhysRevD.106.075003} {\bibfield
  {journal} {\bibinfo  {journal} {Phys. Rev. D}\ }\textbf {\bibinfo {volume}
  {106}},\ \bibinfo {pages} {075003} (\bibinfo {year} {2022})}\BibitemShut
  {NoStop}%
\bibitem [{\citenamefont {Benbrik}\ \emph {et~al.}(2024)\citenamefont
  {Benbrik}, \citenamefont {Boukidi},\ and\ \citenamefont
  {Moretti}}]{PhysRevD.109.055016}%
  \BibitemOpen
  \bibfield  {author} {\bibinfo {author} {\bibfnamefont {R.}~\bibnamefont
  {Benbrik}}, \bibinfo {author} {\bibfnamefont {M.}~\bibnamefont {Boukidi}},\
  and\ \bibinfo {author} {\bibfnamefont {S.}~\bibnamefont {Moretti}},\
  }\bibfield  {title} {\bibinfo {title} {Probing charged higgs bosons in the
  two-higgs-doublet model type ii with vectorlike quarks},\ }\href
  {https://doi.org/10.1103/PhysRevD.109.055016} {\bibfield  {journal} {\bibinfo
   {journal} {Phys. Rev. D}\ }\textbf {\bibinfo {volume} {109}},\ \bibinfo
  {pages} {055016} (\bibinfo {year} {2024})}\BibitemShut {NoStop}%
\bibitem [{\citenamefont {Ouazghour}\ \emph {et~al.}(2019)\citenamefont
  {Ouazghour}, \citenamefont {Arhrib}, \citenamefont {Benbrik}, \citenamefont
  {Chabab},\ and\ \citenamefont {Rahili}}]{PhysRevD.100.035031}%
  \BibitemOpen
  \bibfield  {author} {\bibinfo {author} {\bibfnamefont {B.~A.}\ \bibnamefont
  {Ouazghour}}, \bibinfo {author} {\bibfnamefont {A.}~\bibnamefont {Arhrib}},
  \bibinfo {author} {\bibfnamefont {R.}~\bibnamefont {Benbrik}}, \bibinfo
  {author} {\bibfnamefont {M.}~\bibnamefont {Chabab}},\ and\ \bibinfo {author}
  {\bibfnamefont {L.}~\bibnamefont {Rahili}},\ }\bibfield  {title} {\bibinfo
  {title} {Theory and phenomenology of a two-higgs-doublet type-ii seesaw model
  at the lhc run 2},\ }\href {https://doi.org/10.1103/PhysRevD.100.035031}
  {\bibfield  {journal} {\bibinfo  {journal} {Phys. Rev. D}\ }\textbf {\bibinfo
  {volume} {100}},\ \bibinfo {pages} {035031} (\bibinfo {year}
  {2019})}\BibitemShut {NoStop}%
\bibitem [{\citenamefont {Logan}\ and\ \citenamefont
  {MacLennan}(2010)}]{Logan:2010ag}%
  \BibitemOpen
  \bibfield  {author} {\bibinfo {author} {\bibfnamefont {H.~E.}\ \bibnamefont
  {Logan}}\ and\ \bibinfo {author} {\bibfnamefont {D.}~\bibnamefont
  {MacLennan}},\ }\bibfield  {title} {\bibinfo {title} {{Charged Higgs
  phenomenology in the flipped two Higgs doublet model}},\ }\href
  {https://doi.org/10.1103/PhysRevD.81.075016} {\bibfield  {journal} {\bibinfo
  {journal} {Phys. Rev. D}\ }\textbf {\bibinfo {volume} {81}},\ \bibinfo
  {pages} {075016} (\bibinfo {year} {2010})},\ \Eprint
  {https://arxiv.org/abs/1002.4916} {arXiv:1002.4916 [hep-ph]} \BibitemShut
  {NoStop}%
\bibitem [{\citenamefont {Cordero}\ \emph {et~al.}(2018)\citenamefont
  {Cordero}, \citenamefont {Hernandez-Sanchez}, \citenamefont {Keus},
  \citenamefont {King}, \citenamefont {Moretti}, \citenamefont {Rojas},\ and\
  \citenamefont {Sokolowska}}]{Cordero:2017owj}%
  \BibitemOpen
  \bibfield  {author} {\bibinfo {author} {\bibfnamefont {A.}~\bibnamefont
  {Cordero}}, \bibinfo {author} {\bibfnamefont {J.}~\bibnamefont
  {Hernandez-Sanchez}}, \bibinfo {author} {\bibfnamefont {V.}~\bibnamefont
  {Keus}}, \bibinfo {author} {\bibfnamefont {S.~F.}\ \bibnamefont {King}},
  \bibinfo {author} {\bibfnamefont {S.}~\bibnamefont {Moretti}}, \bibinfo
  {author} {\bibfnamefont {D.}~\bibnamefont {Rojas}},\ and\ \bibinfo {author}
  {\bibfnamefont {D.}~\bibnamefont {Sokolowska}},\ }\bibfield  {title}
  {\bibinfo {title} {{Dark Matter Signals at the LHC from a 3HDM}},\ }\href
  {https://doi.org/10.1007/JHEP05(2018)030} {\bibfield  {journal} {\bibinfo
  {journal} {JHEP}\ }\textbf {\bibinfo {volume} {05}},\ \bibinfo {pages}
  {030}},\ \Eprint {https://arxiv.org/abs/1712.09598} {arXiv:1712.09598
  [hep-ph]} \BibitemShut {NoStop}%
\bibitem [{\citenamefont {Ivanov}\ and\ \citenamefont
  {Vdovin}(2012)}]{Ivanov:2012ry}%
  \BibitemOpen
  \bibfield  {author} {\bibinfo {author} {\bibfnamefont {I.~P.}\ \bibnamefont
  {Ivanov}}\ and\ \bibinfo {author} {\bibfnamefont {E.}~\bibnamefont
  {Vdovin}},\ }\bibfield  {title} {\bibinfo {title} {{Discrete symmetries in
  the three-Higgs-doublet model}},\ }\href
  {https://doi.org/10.1103/PhysRevD.86.095030} {\bibfield  {journal} {\bibinfo
  {journal} {Phys. Rev. D}\ }\textbf {\bibinfo {volume} {86}},\ \bibinfo
  {pages} {095030} (\bibinfo {year} {2012})},\ \Eprint
  {https://arxiv.org/abs/1206.7108} {arXiv:1206.7108 [hep-ph]} \BibitemShut
  {NoStop}%
\bibitem [{\citenamefont {Akeroyd}\ \emph
  {et~al.}(2021{\natexlab{b}})\citenamefont {Akeroyd}, \citenamefont {Logan},
  \citenamefont {Moretti}, \citenamefont {Rojas-Ciofalo}, \citenamefont
  {Shindou},\ and\ \citenamefont {Song}}]{Akeroyd:2021fpf}%
  \BibitemOpen
  \bibfield  {author} {\bibinfo {author} {\bibfnamefont {A.~G.}\ \bibnamefont
  {Akeroyd}}, \bibinfo {author} {\bibfnamefont {H.~E.}\ \bibnamefont {Logan}},
  \bibinfo {author} {\bibfnamefont {S.}~\bibnamefont {Moretti}}, \bibinfo
  {author} {\bibfnamefont {D.}~\bibnamefont {Rojas-Ciofalo}}, \bibinfo {author}
  {\bibfnamefont {T.}~\bibnamefont {Shindou}},\ and\ \bibinfo {author}
  {\bibfnamefont {M.}~\bibnamefont {Song}},\ }\bibfield  {title} {\bibinfo
  {title} {{CP-Violation in the 3-Higgs Doublet Model: CP-Asymmetries from
  Charged Higgs Bosons and Electric Dipole Moments}},\ }\href@noop {} {\
  (\bibinfo {year} {2021}{\natexlab{b}})},\ \Eprint
  {https://arxiv.org/abs/2111.11931} {arXiv:2111.11931 [hep-ph]} \BibitemShut
  {NoStop}%
\bibitem [{\citenamefont {Boto}\ \emph {et~al.}(2025)\citenamefont {Boto},
  \citenamefont {Matos}, \citenamefont {Rom{\~a}o},\ and\ \citenamefont
  {Silva}}]{Boto:2025ovp}%
  \BibitemOpen
  \bibfield  {author} {\bibinfo {author} {\bibfnamefont {R.}~\bibnamefont
  {Boto}}, \bibinfo {author} {\bibfnamefont {J.~A.~C.}\ \bibnamefont {Matos}},
  \bibinfo {author} {\bibfnamefont {J.~C.}\ \bibnamefont {Rom{\~a}o}},\ and\
  \bibinfo {author} {\bibfnamefont {J.~P.}\ \bibnamefont {Silva}},\ }\bibfield
  {title} {\bibinfo {title} {{Surveying the complex three Higgs doublet model
  with Machine Learning}},\ }\href@noop {} {\  (\bibinfo {year} {2025})},\
  \Eprint {https://arxiv.org/abs/2510.02445} {arXiv:2510.02445 [hep-ph]}
  \BibitemShut {NoStop}%
\bibitem [{\citenamefont {Ivanov}\ and\ \citenamefont
  {Nishi}(2015)}]{Ivanov:2014doa}%
  \BibitemOpen
  \bibfield  {author} {\bibinfo {author} {\bibfnamefont {I.~P.}\ \bibnamefont
  {Ivanov}}\ and\ \bibinfo {author} {\bibfnamefont {C.~C.}\ \bibnamefont
  {Nishi}},\ }\bibfield  {title} {\bibinfo {title} {{Symmetry breaking patterns
  in 3HDM}},\ }\href {https://doi.org/10.1007/JHEP01(2015)021} {\bibfield
  {journal} {\bibinfo  {journal} {JHEP}\ }\textbf {\bibinfo {volume} {01}},\
  \bibinfo {pages} {021}},\ \Eprint {https://arxiv.org/abs/1410.6139}
  {arXiv:1410.6139 [hep-ph]} \BibitemShut {NoStop}%
\bibitem [{\citenamefont {Keus}\ \emph {et~al.}(2014)\citenamefont {Keus},
  \citenamefont {King},\ and\ \citenamefont {Moretti}}]{Keus:2013hya}%
  \BibitemOpen
  \bibfield  {author} {\bibinfo {author} {\bibfnamefont {V.}~\bibnamefont
  {Keus}}, \bibinfo {author} {\bibfnamefont {S.~F.}\ \bibnamefont {King}},\
  and\ \bibinfo {author} {\bibfnamefont {S.}~\bibnamefont {Moretti}},\
  }\bibfield  {title} {\bibinfo {title} {{Three-Higgs-doublet models:
  symmetries, potentials and Higgs boson masses}},\ }\href
  {https://doi.org/10.1007/JHEP01(2014)052} {\bibfield  {journal} {\bibinfo
  {journal} {JHEP}\ }\textbf {\bibinfo {volume} {01}},\ \bibinfo {pages}
  {052}},\ \Eprint {https://arxiv.org/abs/1310.8253} {arXiv:1310.8253 [hep-ph]}
  \BibitemShut {NoStop}%
\bibitem [{\citenamefont {Batra}\ \emph {et~al.}(2025)\citenamefont {Batra},
  \citenamefont {Coleppa}, \citenamefont {Khanna}, \citenamefont {Rai},\ and\
  \citenamefont {Sarkar}}]{Batra:2025amk}%
  \BibitemOpen
  \bibfield  {author} {\bibinfo {author} {\bibfnamefont {N.}~\bibnamefont
  {Batra}}, \bibinfo {author} {\bibfnamefont {B.}~\bibnamefont {Coleppa}},
  \bibinfo {author} {\bibfnamefont {A.}~\bibnamefont {Khanna}}, \bibinfo
  {author} {\bibfnamefont {S.~K.}\ \bibnamefont {Rai}},\ and\ \bibinfo {author}
  {\bibfnamefont {A.}~\bibnamefont {Sarkar}},\ }\bibfield  {title} {\bibinfo
  {title} {{Constraining the 3HDM Parameter Space}},\ }\href@noop {} {\
  (\bibinfo {year} {2025})},\ \Eprint {https://arxiv.org/abs/2504.07489}
  {arXiv:2504.07489 [hep-ph]} \BibitemShut {NoStop}%
\bibitem [{\citenamefont {Coleppa}\ \emph {et~al.}(2025)\citenamefont
  {Coleppa}, \citenamefont {Khanna},\ and\ \citenamefont
  {Krishna}}]{Coleppa:2025qst}%
  \BibitemOpen
  \bibfield  {author} {\bibinfo {author} {\bibfnamefont {B.}~\bibnamefont
  {Coleppa}}, \bibinfo {author} {\bibfnamefont {A.}~\bibnamefont {Khanna}},\
  and\ \bibinfo {author} {\bibfnamefont {G.~B.}\ \bibnamefont {Krishna}},\
  }\bibfield  {title} {\bibinfo {title} {{3HDM at the ILC}},\ }\href@noop {} {\
   (\bibinfo {year} {2025})},\ \Eprint {https://arxiv.org/abs/2506.24094}
  {arXiv:2506.24094 [hep-ph]} \BibitemShut {NoStop}%
\bibitem [{\citenamefont {Boto}\ \emph {et~al.}(2023)\citenamefont {Boto},
  \citenamefont {Das}, \citenamefont {Lourenco}, \citenamefont {Rom\~ao},\ and\
  \citenamefont {Silva}}]{PhysRevD.108.015020}%
  \BibitemOpen
  \bibfield  {author} {\bibinfo {author} {\bibfnamefont {R.}~\bibnamefont
  {Boto}}, \bibinfo {author} {\bibfnamefont {D.}~\bibnamefont {Das}}, \bibinfo
  {author} {\bibfnamefont {L.}~\bibnamefont {Lourenco}}, \bibinfo {author}
  {\bibfnamefont {J.~C.}\ \bibnamefont {Rom\~ao}},\ and\ \bibinfo {author}
  {\bibfnamefont {J.~P.}\ \bibnamefont {Silva}},\ }\bibfield  {title} {\bibinfo
  {title} {Fingerprinting the type-z three-higgs-doublet models},\ }\href
  {https://doi.org/10.1103/PhysRevD.108.015020} {\bibfield  {journal} {\bibinfo
   {journal} {Phys. Rev. D}\ }\textbf {\bibinfo {volume} {108}},\ \bibinfo
  {pages} {015020} (\bibinfo {year} {2023})}\BibitemShut {NoStop}%
\bibitem [{\citenamefont {Altmannshofer}\ and\ \citenamefont
  {Toner}(2025)}]{PhysRevD.111.075009}%
  \BibitemOpen
  \bibfield  {author} {\bibinfo {author} {\bibfnamefont {W.}~\bibnamefont
  {Altmannshofer}}\ and\ \bibinfo {author} {\bibfnamefont {K.}~\bibnamefont
  {Toner}},\ }\bibfield  {title} {\bibinfo {title} {Flavor constraints in a
  generational three-higgs-doublet model},\ }\href
  {https://doi.org/10.1103/PhysRevD.111.075009} {\bibfield  {journal} {\bibinfo
   {journal} {Phys. Rev. D}\ }\textbf {\bibinfo {volume} {111}},\ \bibinfo
  {pages} {075009} (\bibinfo {year} {2025})}\BibitemShut {NoStop}%
\bibitem [{\citenamefont {Varzielas}\ and\ \citenamefont
  {Ivanov}(2019)}]{PhysRevD.100.015008}%
  \BibitemOpen
  \bibfield  {author} {\bibinfo {author} {\bibfnamefont {I.~d.~M.}\
  \bibnamefont {Varzielas}}\ and\ \bibinfo {author} {\bibfnamefont {I.~P.}\
  \bibnamefont {Ivanov}},\ }\bibfield  {title} {\bibinfo {title} {Recognizing
  symmetries in a 3hdm in a basis-independent way},\ }\href
  {https://doi.org/10.1103/PhysRevD.100.015008} {\bibfield  {journal} {\bibinfo
   {journal} {Phys. Rev. D}\ }\textbf {\bibinfo {volume} {100}},\ \bibinfo
  {pages} {015008} (\bibinfo {year} {2019})}\BibitemShut {NoStop}%
\bibitem [{\citenamefont {Kalinowski}\ \emph {et~al.}(2023)\citenamefont
  {Kalinowski}, \citenamefont {Kotlarski}, \citenamefont {Rebelo},\ and\
  \citenamefont {de~Medeiros~Varzielas}}]{Kalinowski:2021lvw}%
  \BibitemOpen
  \bibfield  {author} {\bibinfo {author} {\bibfnamefont {J.}~\bibnamefont
  {Kalinowski}}, \bibinfo {author} {\bibfnamefont {W.}~\bibnamefont
  {Kotlarski}}, \bibinfo {author} {\bibfnamefont {M.~N.}\ \bibnamefont
  {Rebelo}},\ and\ \bibinfo {author} {\bibfnamefont {I.}~\bibnamefont
  {de~Medeiros~Varzielas}},\ }\bibfield  {title} {\bibinfo {title} {{3HDM with
  {\ensuremath{\Delta}}(27) symmetry and its phenomenological consequences}},\
  }\href {https://doi.org/10.1007/JHEP02(2023)231} {\bibfield  {journal}
  {\bibinfo  {journal} {JHEP}\ }\textbf {\bibinfo {volume} {02}},\ \bibinfo
  {pages} {231}},\ \Eprint {https://arxiv.org/abs/2112.12699} {arXiv:2112.12699
  [hep-ph]} \BibitemShut {NoStop}%
\bibitem [{\citenamefont {Khachatryan}\ \emph {et~al.}(2015)\citenamefont
  {Khachatryan} \emph {et~al.}}]{CMS:2015flt}%
  \BibitemOpen
  \bibfield  {author} {\bibinfo {author} {\bibfnamefont {V.}~\bibnamefont
  {Khachatryan}} \emph {et~al.} (\bibinfo {collaboration} {CMS}),\ }\bibfield
  {title} {\bibinfo {title} {{Search for a pseudoscalar boson decaying into a Z
  boson and the 125 GeV Higgs boson in $\ell^+\ell^-b\overline{b}$ final
  states}},\ }\href {https://doi.org/10.1016/j.physletb.2015.07.010} {\bibfield
   {journal} {\bibinfo  {journal} {Phys. Lett. B}\ }\textbf {\bibinfo {volume}
  {748}},\ \bibinfo {pages} {221} (\bibinfo {year} {2015})},\ \Eprint
  {https://arxiv.org/abs/1504.04710} {arXiv:1504.04710 [hep-ex]} \BibitemShut
  {NoStop}%
\bibitem [{CMS(2014)}]{CMS-PAS-HIG-14-011}%
  \BibitemOpen
  \href {https://cds.cern.ch/record/1969698} {\emph {\bibinfo {title} {{Search
  for a pseudoscalar boson A decaying into a Z and an h boson in the llbb final
  state}}}},\ \bibinfo {type} {Tech. Rep.}\ (\bibinfo  {institution} {CERN},\
  \bibinfo {address} {Geneva},\ \bibinfo {year} {2014})\BibitemShut {NoStop}%
\bibitem [{\citenamefont {Sirunyan}\ \emph {et~al.}(2019)\citenamefont
  {Sirunyan} \emph {et~al.}}]{CMS:2019qcx}%
  \BibitemOpen
  \bibfield  {author} {\bibinfo {author} {\bibfnamefont {A.~M.}\ \bibnamefont
  {Sirunyan}} \emph {et~al.} (\bibinfo {collaboration} {CMS}),\ }\bibfield
  {title} {\bibinfo {title} {{Search for a heavy pseudoscalar boson decaying to
  a Z and a Higgs boson at $\sqrt{s} =$ 13 TeV}},\ }\href
  {https://doi.org/10.1140/epjc/s10052-019-7058-z} {\bibfield  {journal}
  {\bibinfo  {journal} {Eur. Phys. J. C}\ }\textbf {\bibinfo {volume} {79}},\
  \bibinfo {pages} {564} (\bibinfo {year} {2019})},\ \Eprint
  {https://arxiv.org/abs/1903.00941} {arXiv:1903.00941 [hep-ex]} \BibitemShut
  {NoStop}%
\bibitem [{\citenamefont {Bento}\ \emph {et~al.}(2022)\citenamefont {Bento},
  \citenamefont {Rom\~ao},\ and\ \citenamefont {Silva}}]{Bento:2022vsb}%
  \BibitemOpen
  \bibfield  {author} {\bibinfo {author} {\bibfnamefont {M.~P.}\ \bibnamefont
  {Bento}}, \bibinfo {author} {\bibfnamefont {J.~C.}\ \bibnamefont {Rom\~ao}},\
  and\ \bibinfo {author} {\bibfnamefont {J.~a.~P.}\ \bibnamefont {Silva}},\
  }\bibfield  {title} {\bibinfo {title} {{Unitarity bounds for all
  symmetry-constrained 3HDMs}},\ }\href
  {https://doi.org/10.1007/JHEP08(2022)273} {\bibfield  {journal} {\bibinfo
  {journal} {JHEP}\ }\textbf {\bibinfo {volume} {08}},\ \bibinfo {pages}
  {273}},\ \Eprint {https://arxiv.org/abs/2204.13130} {arXiv:2204.13130
  [hep-ph]} \BibitemShut {NoStop}%
\bibitem [{\citenamefont {Bechtle}\ \emph {et~al.}(2020)\citenamefont
  {Bechtle}, \citenamefont {Dercks}, \citenamefont {Heinemeyer}, \citenamefont
  {Klingl}, \citenamefont {Stefaniak}, \citenamefont {Weiglein},\ and\
  \citenamefont {Wittbrodt}}]{Bechtle:2020pkv}%
  \BibitemOpen
  \bibfield  {author} {\bibinfo {author} {\bibfnamefont {P.}~\bibnamefont
  {Bechtle}}, \bibinfo {author} {\bibfnamefont {D.}~\bibnamefont {Dercks}},
  \bibinfo {author} {\bibfnamefont {S.}~\bibnamefont {Heinemeyer}}, \bibinfo
  {author} {\bibfnamefont {T.}~\bibnamefont {Klingl}}, \bibinfo {author}
  {\bibfnamefont {T.}~\bibnamefont {Stefaniak}}, \bibinfo {author}
  {\bibfnamefont {G.}~\bibnamefont {Weiglein}},\ and\ \bibinfo {author}
  {\bibfnamefont {J.}~\bibnamefont {Wittbrodt}},\ }\bibfield  {title} {\bibinfo
  {title} {{HiggsBounds-5: Testing Higgs Sectors in the LHC 13 TeV Era}},\
  }\href {https://doi.org/10.1140/epjc/s10052-020-08557-9} {\bibfield
  {journal} {\bibinfo  {journal} {Eur. Phys. J. C}\ }\textbf {\bibinfo {volume}
  {80}},\ \bibinfo {pages} {1211} (\bibinfo {year} {2020})},\ \Eprint
  {https://arxiv.org/abs/2006.06007} {arXiv:2006.06007 [hep-ph]} \BibitemShut
  {NoStop}%
\bibitem [{\citenamefont {Bahl}\ \emph {et~al.}(2023)\citenamefont {Bahl},
  \citenamefont {Biek\"otter}, \citenamefont {Heinemeyer}, \citenamefont {Li},
  \citenamefont {Paasch}, \citenamefont {Weiglein},\ and\ \citenamefont
  {Wittbrodt}}]{Bahl:2022igd}%
  \BibitemOpen
  \bibfield  {author} {\bibinfo {author} {\bibfnamefont {H.}~\bibnamefont
  {Bahl}}, \bibinfo {author} {\bibfnamefont {T.}~\bibnamefont {Biek\"otter}},
  \bibinfo {author} {\bibfnamefont {S.}~\bibnamefont {Heinemeyer}}, \bibinfo
  {author} {\bibfnamefont {C.}~\bibnamefont {Li}}, \bibinfo {author}
  {\bibfnamefont {S.}~\bibnamefont {Paasch}}, \bibinfo {author} {\bibfnamefont
  {G.}~\bibnamefont {Weiglein}},\ and\ \bibinfo {author} {\bibfnamefont
  {J.}~\bibnamefont {Wittbrodt}},\ }\bibfield  {title} {\bibinfo {title}
  {{HiggsTools: BSM scalar phenomenology with new versions of HiggsBounds and
  HiggsSignals}},\ }\href {https://doi.org/10.1016/j.cpc.2023.108803}
  {\bibfield  {journal} {\bibinfo  {journal} {Comput. Phys. Commun.}\ }\textbf
  {\bibinfo {volume} {291}},\ \bibinfo {pages} {108803} (\bibinfo {year}
  {2023})},\ \Eprint {https://arxiv.org/abs/2210.09332} {arXiv:2210.09332
  [hep-ph]} \BibitemShut {NoStop}%
\bibitem [{\citenamefont {Bechtle}\ \emph {et~al.}(2021)\citenamefont
  {Bechtle}, \citenamefont {Heinemeyer}, \citenamefont {Klingl}, \citenamefont
  {Stefaniak}, \citenamefont {Weiglein},\ and\ \citenamefont
  {Wittbrodt}}]{Bechtle:2020uwn}%
  \BibitemOpen
  \bibfield  {author} {\bibinfo {author} {\bibfnamefont {P.}~\bibnamefont
  {Bechtle}}, \bibinfo {author} {\bibfnamefont {S.}~\bibnamefont {Heinemeyer}},
  \bibinfo {author} {\bibfnamefont {T.}~\bibnamefont {Klingl}}, \bibinfo
  {author} {\bibfnamefont {T.}~\bibnamefont {Stefaniak}}, \bibinfo {author}
  {\bibfnamefont {G.}~\bibnamefont {Weiglein}},\ and\ \bibinfo {author}
  {\bibfnamefont {J.}~\bibnamefont {Wittbrodt}},\ }\bibfield  {title} {\bibinfo
  {title} {{HiggsSignals-2: Probing new physics with precision Higgs
  measurements in the LHC 13 TeV era}},\ }\href
  {https://doi.org/10.1140/epjc/s10052-021-08942-y} {\bibfield  {journal}
  {\bibinfo  {journal} {Eur. Phys. J. C}\ }\textbf {\bibinfo {volume} {81}},\
  \bibinfo {pages} {145} (\bibinfo {year} {2021})},\ \Eprint
  {https://arxiv.org/abs/2012.09197} {arXiv:2012.09197 [hep-ph]} \BibitemShut
  {NoStop}%
\bibitem [{\citenamefont {Borzumati}\ and\ \citenamefont
  {Greub}(1998)}]{Borzumati_1998}%
  \BibitemOpen
  \bibfield  {author} {\bibinfo {author} {\bibfnamefont {F.~M.}\ \bibnamefont
  {Borzumati}}\ and\ \bibinfo {author} {\bibfnamefont {C.}~\bibnamefont
  {Greub}},\ }\bibfield  {title} {\bibinfo {title} {Two higgs doublet model
  predictions for $b \rightarrow x_s \gamma$ in nlo qcd},\ }\bibfield
  {journal} {\bibinfo  {journal} {Physical Review D}\ }\textbf {\bibinfo
  {volume} {58}},\ \href {https://doi.org/10.1103/physrevd.58.074004}
  {10.1103/physrevd.58.074004} (\bibinfo {year} {1998})\BibitemShut {NoStop}%
\bibitem [{\citenamefont {Boto}\ \emph {et~al.}(2021)\citenamefont {Boto},
  \citenamefont {Romão},\ and\ \citenamefont {Silva}}]{Boto_2021}%
  \BibitemOpen
  \bibfield  {author} {\bibinfo {author} {\bibfnamefont {R.}~\bibnamefont
  {Boto}}, \bibinfo {author} {\bibfnamefont {J.~C.}\ \bibnamefont {Romão}},\
  and\ \bibinfo {author} {\bibfnamefont {J.~P.}\ \bibnamefont {Silva}},\
  }\bibfield  {title} {\bibinfo {title} {Current bounds on the type-z $z_3$
  three-higgs-doublet model},\ }\bibfield  {journal} {\bibinfo  {journal}
  {Physical Review D}\ }\textbf {\bibinfo {volume} {104}},\ \href
  {https://doi.org/10.1103/physrevd.104.095006} {10.1103/physrevd.104.095006}
  (\bibinfo {year} {2021})\BibitemShut {NoStop}%
\bibitem [{\citenamefont {Alwall}\ \emph {et~al.}(2011)\citenamefont {Alwall},
  \citenamefont {Herquet}, \citenamefont {Maltoni}, \citenamefont {Mattelaer},\
  and\ \citenamefont {Stelzer}}]{Alwall:2011uj}%
  \BibitemOpen
  \bibfield  {author} {\bibinfo {author} {\bibfnamefont {J.}~\bibnamefont
  {Alwall}}, \bibinfo {author} {\bibfnamefont {M.}~\bibnamefont {Herquet}},
  \bibinfo {author} {\bibfnamefont {F.}~\bibnamefont {Maltoni}}, \bibinfo
  {author} {\bibfnamefont {O.}~\bibnamefont {Mattelaer}},\ and\ \bibinfo
  {author} {\bibfnamefont {T.}~\bibnamefont {Stelzer}},\ }\bibfield  {title}
  {\bibinfo {title} {{MadGraph 5 : Going Beyond}},\ }\href
  {https://doi.org/10.1007/JHEP06(2011)128} {\bibfield  {journal} {\bibinfo
  {journal} {JHEP}\ }\textbf {\bibinfo {volume} {06}},\ \bibinfo {pages}
  {128}},\ \Eprint {https://arxiv.org/abs/1106.0522} {arXiv:1106.0522 [hep-ph]}
  \BibitemShut {NoStop}%
\bibitem [{\citenamefont {Staub}(2008)}]{Staub:2008uz}%
  \BibitemOpen
  \bibfield  {author} {\bibinfo {author} {\bibfnamefont {F.}~\bibnamefont
  {Staub}},\ }\bibfield  {title} {\bibinfo {title} {{SARAH}},\ }\href@noop {}
  {\  (\bibinfo {year} {2008})},\ \Eprint {https://arxiv.org/abs/0806.0538}
  {arXiv:0806.0538 [hep-ph]} \BibitemShut {NoStop}%
\bibitem [{\citenamefont {Porod}(2003)}]{Porod:2003um}%
  \BibitemOpen
  \bibfield  {author} {\bibinfo {author} {\bibfnamefont {W.}~\bibnamefont
  {Porod}},\ }\bibfield  {title} {\bibinfo {title} {{SPheno, a program for
  calculating supersymmetric spectra, SUSY particle decays and SUSY particle
  production at e+ e- colliders}},\ }\href
  {https://doi.org/10.1016/S0010-4655(03)00222-4} {\bibfield  {journal}
  {\bibinfo  {journal} {Comput. Phys. Commun.}\ }\textbf {\bibinfo {volume}
  {153}},\ \bibinfo {pages} {275} (\bibinfo {year} {2003})},\ \Eprint
  {https://arxiv.org/abs/hep-ph/0301101} {arXiv:hep-ph/0301101} \BibitemShut
  {NoStop}%
\bibitem [{\citenamefont {Bierlich}\ \emph {et~al.}(2022)\citenamefont
  {Bierlich} \emph {et~al.}}]{Bierlich:2022pfr}%
  \BibitemOpen
  \bibfield  {author} {\bibinfo {author} {\bibfnamefont {C.}~\bibnamefont
  {Bierlich}} \emph {et~al.},\ }\bibfield  {title} {\bibinfo {title} {{A
  comprehensive guide to the physics and usage of PYTHIA 8.3}},\ }\href
  {https://doi.org/10.21468/SciPostPhysCodeb.8} {\bibfield  {journal} {\bibinfo
   {journal} {SciPost Phys. Codeb.}\ }\textbf {\bibinfo {volume} {2022}},\
  \bibinfo {pages} {8} (\bibinfo {year} {2022})},\ \Eprint
  {https://arxiv.org/abs/2203.11601} {arXiv:2203.11601 [hep-ph]} \BibitemShut
  {NoStop}%
\bibitem [{\citenamefont {de~Favereau}\ \emph {et~al.}(2014)\citenamefont
  {de~Favereau}, \citenamefont {Delaere}, \citenamefont {Demin}, \citenamefont
  {Giammanco}, \citenamefont {Lema{\^\i}tre}, \citenamefont {Mertens},\ and\
  \citenamefont {Selvaggi}}]{deFavereau:2013fsa}%
  \BibitemOpen
  \bibfield  {author} {\bibinfo {author} {\bibfnamefont {J.}~\bibnamefont
  {de~Favereau}}, \bibinfo {author} {\bibfnamefont {C.}~\bibnamefont
  {Delaere}}, \bibinfo {author} {\bibfnamefont {P.}~\bibnamefont {Demin}},
  \bibinfo {author} {\bibfnamefont {A.}~\bibnamefont {Giammanco}}, \bibinfo
  {author} {\bibfnamefont {V.}~\bibnamefont {Lema{\^\i}tre}}, \bibinfo {author}
  {\bibfnamefont {A.}~\bibnamefont {Mertens}},\ and\ \bibinfo {author}
  {\bibfnamefont {M.}~\bibnamefont {Selvaggi}} (\bibinfo {collaboration}
  {DELPHES 3}),\ }\bibfield  {title} {\bibinfo {title} {{DELPHES 3, A modular
  framework for fast simulation of a generic collider experiment}},\ }\href
  {https://doi.org/10.1007/JHEP02(2014)057} {\bibfield  {journal} {\bibinfo
  {journal} {JHEP}\ }\textbf {\bibinfo {volume} {02}},\ \bibinfo {pages}
  {057}},\ \Eprint {https://arxiv.org/abs/1307.6346} {arXiv:1307.6346 [hep-ex]}
  \BibitemShut {NoStop}%
\bibitem [{\citenamefont {Conte}\ \emph {et~al.}(2013)\citenamefont {Conte},
  \citenamefont {Fuks},\ and\ \citenamefont {Serret}}]{Conte:2012fm}%
  \BibitemOpen
  \bibfield  {author} {\bibinfo {author} {\bibfnamefont {E.}~\bibnamefont
  {Conte}}, \bibinfo {author} {\bibfnamefont {B.}~\bibnamefont {Fuks}},\ and\
  \bibinfo {author} {\bibfnamefont {G.}~\bibnamefont {Serret}},\ }\bibfield
  {title} {\bibinfo {title} {{MadAnalysis 5, A User-Friendly Framework for
  Collider Phenomenology}},\ }\href {https://doi.org/10.1016/j.cpc.2012.09.009}
  {\bibfield  {journal} {\bibinfo  {journal} {Comput. Phys. Commun.}\ }\textbf
  {\bibinfo {volume} {184}},\ \bibinfo {pages} {222} (\bibinfo {year}
  {2013})},\ \Eprint {https://arxiv.org/abs/1206.1599} {arXiv:1206.1599
  [hep-ph]} \BibitemShut {NoStop}%
\end{thebibliography}%

\end{document}